\documentclass{mn2e}
\usepackage{graphicx,times}

\def \kms {${\rm{km}\,\rm{s}^{-1}}$}

\def \apjs{ApJS}
\def \aaps{A\&AS}
\def \aap{A\&A}
\def \apj{ApJ}
\def \aj{AJ}
\def \mnras{MNRAS}

\title[Gaseous stripping candidates in Coma]{Ultraviolet tails and trails in cluster galaxies:\\A sample of candidate gaseous stripping events in Coma}

\author[Russell J. Smith et al. ]
{Russell J. Smith$^{1}$\thanks{Email: russell.smith@durham.ac.uk}, 
John R. Lucey$^{1}$, 
Derek Hammer$^{2,3}$, 
Ann E. Hornschemeier$^{3,2}$,
\newauthor
David Carter$^{4}$, 
Michael J. Hudson$^{5,6,7}$, 
Ronald O. Marzke$^{8}$, 
Mustapha Mouhcine$^{4}$,
\newauthor
Sareh Eftekharzadeh$^{9}$, 
Phil James$^{4}$, 
Habib Khosroshahi$^{9}$, 
Ehsan Kourkchi$^{9}$, 
Arna Karick$^{4}$
\\
~\\
$^1$Department of Physics, University of Durham, Durham DH1 3LE\\
$^2$Department of Physics and Astronomy, Johns Hopkins University, 3400 North Charles Street, Baltimore, MD 21218, USA\\
$^3$NASA Goddard Space Flight Centre, Code 662.0, Greenbelt, MD 20771, USA\\
$^4$Astronomical Research Institute, Liverpool John Moores University, Twelve Quays House, Egerton Wharf , Birkenhead CH41 1LD\\
$^5$Department of Physics and Astronomy, University of Waterloo, 200 University Avenue West, Waterloo, Ontario N2L 3G1, Canada\\
$^6$Institut d'Astrophysique de Paris, UMR 7095 / Universit\'e Pierre et Marie Curie, 98bis boulevard Arago, 75014, Paris, France\\
$^7$Perimeter Institute for Theoretical Physics, 31 Caroline St. N., Waterloo, ON, N2L 2Y5, Canada\\
$^8$Department of Physics and Astronomy, San Francisco State University, San Francisco, CA 94132, USA\\
$^9$School of Astronomy, Institute for Research in Fundamental Sciences, PO Box 19395-5531, Tehran, Iran
}

\pagerange{\pageref{firstpage}$-$\pageref{lastpage}} \pubyear{2010}

\begin{document}

\label{firstpage}

\maketitle

\begin{abstract}
We have used new deep observations of the Coma cluster from {\it Galaxy Evolution Explorer} to visually identify 13 star-forming galaxies
with asymmetric morphologies in the ultraviolet. Aided by wide-field optical broad-band and H$\alpha$ imaging, we interpret the asymmetric 
features as being due to star formation within gas stripped from the galaxies by interaction with the cluster environment. 
The selected objects display a range of structures from broad fan-shaped systems of filaments and knots (``jellyfish") to narrower and smoother tails 
extending up to 100\,kpc in length. Some of the features have been discussed previously in the literature, while others are newly identified here. 
We assess the ensemble properties of the sample. The candidate stripping events are located closer to the cluster centre than other star-forming galaxies; 
their radial distribution is more similar to that of all cluster members, dominated by passive galaxies. The fraction of blue galaxies which 
are undergoing stripping falls from 40\,per cent in the central 500\,kpc, to less than 5\,per cent beyond 1\,Mpc.
We find that tails pointing away from (i.e. galaxies moving towards) the cluster centre are strongly favoured (11/13 cases). 
From the small number of ``outgoing'' galaxies with stripping signatures, we conclude that the stripping events occur primarily on first passage 
towards the cluster centre, and are short-lived compared to the cluster crossing time. 
Using galaxy infall trajectories extracted from a cosmological simulation, we find that the observed fraction of 
blue galaxies undergoing stripping can be reproduced if the events are triggered at a threshold radius of $\sim$1\,Mpc
and detectable for $\sim$500\,Myr. 
{\it Hubble Space Telescope} images are available for two galaxies from our sample and reveal compact blue knots coincident with 
UV and H$\alpha$ emission, apparently forming stars within the stripped material. 
Our results confirm that stripping of gas from infalling galaxies, and associated star formation in the stripped material, is a widespread 
phenomenon in rich clusters. Deep UV imaging of additional clusters is a promising route to constructing a statistically powerful sample
of stripping events and constraining models for the truncation of star formation in clusters.
\end{abstract}


\section{Introduction}

In rich galaxy clusters, the evolution of the member galaxies and the intra-cluster medium (ICM) are intricately linked, through various forms 
of ejection of material from galaxies into their surroundings. Galaxies in dense environments are susceptible to loss of stellar mass through 
tidal stripping, 
removal of cold gas from their disks by ram-pressure stripping (e.g. Gunn \& Gott 1972; Quilis, Moore \& Bower 2000), 
and stripping of hot gas from their halos by the same mechanism (e.g. Larson, Tinsley \& Caldwell 1980; McCarthy et al. 2008).
The eventual consequences are as varied as the
physical processes involved: depletion of cold gas truncates star formation the disks, perhaps leading to the formation of cluster S0 galaxies; 
halo-stripping causes a more gradual decline in the star-formation history (``strangulation'' or ``starvation''); 
tidal stripping transfers individual stars and globular clusters to the intra-cluster population (e.g. Peng et al. 2010), and may disrupt
dwarfs enough to affect the luminosity function (e.g. Henriques, Bertone \& Thomas 2008); stripped cold gas may be heated and mixed with the ICM, 
or survive in clumps that can heat the cluster core (Dekel \& Birnboim 2008), or fragment to form new stars or clusters in the intra-cluster
population  (Yoshida et al. 2008; Puchwein et al. 2010). 

There is substantial observational evidence for gaseous (as opposed to purely stellar) stripping events in clusters. 
Notable examples are seen in Virgo (e.g. Phookun \& Mundy 1995; Crowl \& Kenney 2006; Vollmer et al. 2009), 
Abell 3627 (Sun, Donahue \& Voit 2007; Woudt et al. 2008; Sun et al. 2010), 
Abell 1367 (e.g. Gavazzi et al. 2001; Scott et al. 2010), as well as in Coma (e.g. Vollmer et al. 2001; Yagi et al. 2007; Yoshida et al. 2008).
Two particularly spectacular cases have been highlighted by Cortese et al. (2007) in Abell 1689 and Abell 2667 at $z\sim0.2$. 
Another example in a distant cluster (Abell 2125 at $z=0.25$) is discussed by Owen et al. (2006).  

Simple calculations based on the arguments of Gunn \& Gott (1972) suggest that ram-pressure stripping is  
effective mainly in the cores of clusters. Several groups have recently performed detailed simulations of individual galaxies
to investigate the properties of the stripped tails (e.g. Roediger \& Br\"uggen 2008; Kapferer et al. 2009; Tonnesen \& Bryan 2010). 
The simulations with radiative cooling (Kapferer et al.;Tonnesen \& Bryan) show narrow, highly structured
tails, with dense clouds embedded within them. When star formation is included (Kapferer et al.), the stars created in these 
dense clouds in some cases fall back towards the stripped galaxy, adding to its bulge. 

Until now, observational studies of ongoing stripping in clusters were based on one or two galaxies per cluster, in clusters of 
disparate properties and a range in redshift. As a result, meaningful statistical conclusions about the stripped galaxies could not be drawn. 
In this paper, we use wide-field UV and optical imaging data to identify a {\it sample} of 
candidate gaseous stripping events in a single nearby cluster (Coma), primarily on the basis of emission from young stars formed in the stripped gas. 
Rather than concentrate on details of the individual cases, we instead focus on the ensemble statistics of   
galaxies undergoing this process, including their radial and redshift
distributions compared to other cluster galaxies, and the orientation of their projected velocities relative to the cluster centre. 

The paper is structured as follows:
The observations and selection of candidate stripping events are described in Section~\ref{sec:obs}, followed by a brief description of each of the
selected objects, with reference to previous work. The spatial and redshift distribution of the sample is analysed in Section~\ref{sec:stats}
and interpreted in Section~\ref{sec:disc}. Our conclusions are summarized in Section~\ref{sec:concs}. 

Throughout the paper, we adopt a distance 
of 100\,Mpc for Coma, so that the distance modulus is 35.0 and one arcminute corresponds to  29\,kpc.

\section{Sample construction}\label{sec:obs}

\subsection{Observations}

Our primary observational resources are deep ultraviolet (UV) imaging from the {\it Galaxy Evolution Explorer} ({\it {\it GALEX}}) satellite, and 
$ugi$ optical imaging from the 3.6m Canada--France--Hawaii Telescope (CFHT). 
The {\it GALEX} data are especially useful to search for stripping events because the formation of young stars in the 
stripped material leads to high contrast in the UV. Accordingly, we limit our study to the area covered by two deep {\it GALEX} observations. 

In the core of the cluster, we obtained a deep observation in {\it GALEX} Cycle 5 (PI: Smith), in `petal-pattern' mode to ensure
detector safety due to the presence of UV-bright stars in the field. The total exposure time in the co-added data used for our
analysis is 14.7\,ksec in the NUV detector and 13.6\,ksec in the FUV.
In the south-west part of the cluster, we use the deep observation
made in {\it GALEX} Cycle 2 (PI: Hornschemeier), and analysed by Hammer et al. (2010a). The co-added images used for the present analysis,
have a depth of 21.0/19.8\,ksec (NUV/FUV), slightly less than the total integration used by Hammer et al. 
Together, the {\it GALEX} fields extend nearly to the virial radius of the cluster, and include the well-known infalling group centred on NGC 4839
(Briel, Henry \& B\"ohringer 1992). 
In the core field, the surface brightness limit for detecting diffuse structures (which we define as a 3$\sigma$ fluctuation on 10$\times$10\,arcsec areas) is 
29.7\,mag\,arcsec$^{-2}$ in NUV and 30.1\,mag\,arcsec$^{-2}$ in FUV. 

Complementary optical data have been obtained with MegaCam at the CFHT (PI: Hudson), 
imaging a 3$\times$3\,deg$^2$ area centred on the cluster core. The exposure times were 1360s, 300s, 300s in $u$, $g$ and $i$ bands. 
After pipeline processing and stacking, the 80\,per cent completeness
depths for point sources are at least 24.5, 24.0 and 23.5 in $u$, $g$ and $i$, with variations from 
field to field due to varying sky brightness. In the central field, the surface brightness limit (defined as above) is 27.6, 27.4 and 26.5
mag\,arcsec$^{-2}$ in $u$, $g$ and $i$.
The image quality in the $i$-band ranges from 0.5\,arcsec to 0.8\,arcsec FWHM, with a median of  $\sim$0.6\,arcsec. 

Within the central part of the cluster, we also employed a much deeper $u$-band MegaCam dataset obtained
from the archive (PI: Adami), using a custom stack kindly generated by Dr S. Gwyn, with a total integration of 23400s, 
and FWHM 1.3\,arcsec. The surface brightness limit is 28.5\,mag\,arcsec$^{-2}$.
This image was not used for construction of catalogues, but only as an additional visual
reference for faint features in the $u$-band. 

We made similar incidental use of H$\alpha$ imaging data from the 2.5m Isaac Newton Telescope (INT, PI: Mouhcine). The H$\alpha$ data 
are part of a wide-field survey covering $\sim$2.5\,deg$^2$ of the cluster and sensitive to H$\alpha$ emission in the redshift range
$cz=4900-10800$\,\kms. 
These data will be presented and analysed elsewhere. Here, we used
continuum-subtracted H$\alpha$ images of the {\it GALEX}-selected candidate stripping events, 
to assess the distribution of ongoing star formation in the stripped systems. 

\begin{figure}
\includegraphics[angle=270,width=85mm]{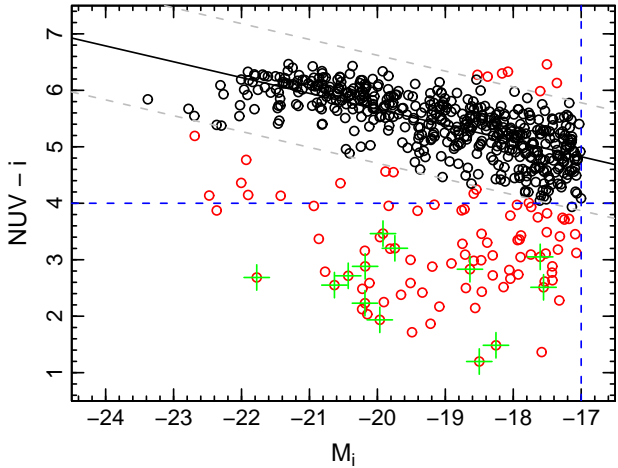}
\caption{The UV versus optical colour--magnitude relation for spectroscopically-confirmed cluster members within the two {\it GALEX}
fields. No galactic extinction corrections or $k$-corrections have been applied.
The galaxies plotted in red are outliers from a robust fit to the red sequence, indicated by the solid line. 
Objects bluer than the dashed line at $NUV-i=4$ form the sample examined for signs of gaseous stripping. 
The 13 candidate stripping events identified in this paper are highlighted by green crosses.}
\label{fig:cmr}
\end{figure}

\subsection{Catalogues}

In order to measure matched-aperture colours, the MegaCam images were resampled,  using {\sc swarp} (Bertin et al. 2002),
to the same 1.5-arcsec pixels as the {\it GALEX} data, and smoothed to match the $\sim$5\,arcsec FWHM point-spread 
function (PSF) of the UV images. In the latter step, we applied a circular gaussian PSF, constant over all of the data. Although
variation and ellipticity of the PSF should be taken into account for photometry of higher precision, these approximations
are justified for our purposes.
 
Photometry was performed using {\sc SExtractor} (Bertin \& Arnouts 1996) in dual-image mode. The MegaCam $i$-band images were used for object detection, 
deblending and aperture definition. Magnitudes for the detected objects were then computed within common apertures
for the two {\it GALEX} and three MegaCam bandpasses. In this paper, all magnitudes and colours are based on counts within
the $i$-band elliptical Kron-like aperture.

Finally, the catalogues were matched to a compilation of redshift data, including a deep survey made with Hectospec at the 6.5m MMT 
(Marzke et al. in preparation). Within the surveyed fields, there are 589 known member galaxies, defined by 3000\,\kms$< cz <$13000\,\kms, having
$i\le18$ ($M_i\le-17$), and all but two of these were recovered in the {\it GALEX} data with positive flux. Figure~\ref{fig:cmr} shows the $NUV-i$ 
colour--magnitude relation for the resulting
galaxy sample. A clear red sequence of passive galaxies is seen, with a scatter of 0.38\,mag around a linear fit, which is typical for the NUV colours 
(e.g. Rawle et al. 2008). The systematic deviation from the linear fit for the brightest galaxies is likely due to contamination from the UV-excess 
sources (old hot stars) which dominate at FUV but also contribute substantial flux at NUV. 

\begin{table*}
\caption{Identifications, co-ordinates and other data for the galaxies discussed in this paper. GMP numbers are from Godwin, Metcalfe \& Peach (1983).
P.A. is the average position angle (in degrees east of north), of the stripped material estimated by eye from the images. 
The angle $\theta_{\rm \ \ clus}$ is measured between the tail and the cluster-centric vector, such that
$\theta_{\rm \ clus}=0^\circ$ is a tail directed away from the cluster centre, and $\theta_{\rm \ clus}=180^\circ$ 
is a tail directed towards the cluster centre. The column headed ``type'' gives the morphological class assigned by Dressler (1980).
H$\alpha$ indicates whether extended or asymmetric H$\alpha$ emission is observed in our INT data. The ``comp'' column indicates
whether the galaxy has a bright projected neighbour galaxy with a small radial velocity difference ($<500$\,\kms).
The final column is the leading identification in the NASA Extragalactic Database. 
}
\label{tab:gsesample}
\begin{tabular}{lrrrcccrrcccl}
\hline
ID & \multicolumn{1}{c}{R.A.} & \multicolumn{1}{c}{Dec} & \multicolumn{1}{c}{$D_{\rm cl}$} & \multicolumn{1}{c}{$cz$}  & \multicolumn{1}{c}{$i$} & \multicolumn{1}{c}{$NUV-i$} & \multicolumn{1}{c}{P.A.} &  \multicolumn{1}{c}{$\theta_{\rm clus}$} & \multicolumn{1}{c}{Type} & H$\alpha$ & comp & NED name \\
 &  \multicolumn{1}{c}{(J2000)} & \multicolumn{1}{c}{(J2000)} &  \multicolumn{1}{c}{[kpc]} & [km\,s$^{-1}$] & [mag] & [mag] & [deg] & [deg] & & & & \\
\hline
GMP 2559 & 195.158 & 28.057 & 448 & 7845 & 14.37 & 2.55 & 177 & 110 & Scd  &y&n& IC 4040 \\ 
GMP 2599 & 195.140 & 27.638 & 694 & 7497 & 15.04 & 1.94 & 141 & 5 &  Sb &y&n& KUG 1258+279A \\ 
GMP 2640 & 195.122 & 27.515 & 874 & 7429 & 15.26 & 3.20 & 336 & 179 & S0p  &n&y& KUG 1258+277 \\ 
GMP 2910 & 195.038 & 27.866 & 277 & 5297 & 15.08 & 3.46 & 62 & 65 &  I &y&n& MRK 0060 NED01 \\ 
GMP 3016 & 195.004 & 28.082 & 278 & 7765 & 17.40 & 3.05 & 40 & 3 & ---  &n&n& MAPS-NGP 0$\_$323$\_$0924669 \\ 
GMP 3816 & 194.759 & 28.116 & 360 & 9431 & 14.82 & 2.23 & 326 & 4 & Sbc  &y&n& NGC 4858 \\ 
GMP 4060 & 194.678 & 27.760 & 501 & 8756 & 16.36 & 2.83 & 180 & 45 & --- &y&n& SDSS J125842.58+274537.8 \\ 
GMP 4232 & 194.628 & 27.564 & 831 & 7283 & 17.45 & 2.51 & 138 & 73 & --- &y&y& MAPS-NGP 0$\_$323$\_$1033883 \\ 
GMP 4471 & 194.523 & 28.243 & 786 & 7193 & 13.22 & 2.69 & 332 & 21 & Scd &y&n& NGC 4848 \\ 
GMP 4555 & 194.491 & 28.061 & 675 & 8163 & 14.82 & 2.88 & 235 & 51 & I &y&n& KUG 1255+283 \\ 
GMP 4570 & 194.486 & 27.992 & 659 & 4595 & 16.50 & 1.20 & 304 & 29 & I &n&n& SDSS J125756.79+275930.3 \\ 
GMP 4629 & 194.459 & 28.170 & 796 & 6936 & 16.75 & 1.48 & 344 & 46 & --- &y&y& SDSS J125750.23+281013.2 \\ 
GMP 5422 & 194.119 & 27.291 & 1727 & 7532 & 14.57 & 2.72 & 246 & 20 &  --- &---&n& IC 3913 \\ 
\hline
\end{tabular}
\end{table*}

\subsection{Visual selection of candidate gaseous stripping events}

We aim to identify stripping events that are due to ``gaseous'' interaction with the surrounding medium. Under this definition, we
include several processes including ram-pressure stripping, viscous stripping, and also tidal interactions perturbing the gas. 
However, we do not include tidal disturbance or disruption of {\it purely} stellar systems, which may also be common in clusters
(e.g. Gregg \& West 1998), but which are not sensitively probed by UV data. 
Although gaseous interactions directly perturb the gas distribution (which can be observed in HI 21\,cm emission), they may also lead to 
the formation of new stars within the stripped material which strongly affect the appearance of the galaxy in the UV
regime. Asymmetric features in the UV morphology thus provide an {\it indirect} probe of ongoing (or at least very recent) gaseous stripping. 
The primary criterion we apply to select a sample of such events is therefore UV asymmetry visible in the {\it GALEX} image, 
without a strong corresponding disturbance in the redder bands of the MegaCam data. 

An initial visual search in the central {\it GALEX} field revealed a number of candidates for recent gaseous stripping, including
several previously noted in the literature. 
All have blue UV-to-optical colours, with $NUV-i < 4$, compared to red-sequence colours
of $5 < (NUV-i) < 6$. To construct a more controlled sample, all 80 galaxies with  $NUV-i < 4$ and $M_i<-17$
(see Figure~\ref{fig:cmr}) were subsequently examined closely in the UV and optical images, to search for comparable features. 
The simple constant colour threshold was adopted for simplicity; examining the galaxies that would be selected by a sloping cut that 
follows the red sequence, no additional similar objects were found.
During the visual examination of blue galaxies, ten were judged to be due to blends with neighbouring galaxies or contaminated by
image edges, ghosts, halos from bright stars, etc. These are removed from the sample of blue objects when computing fractions
and distributions in the following sections. 

We identified 13 galaxies showing UV asymmetry which we interpret as evidence for ongoing gaseous stripping
(Table~\ref{tab:gsesample} and Figure~\ref{fig:images2}).
Some galaxies show clearly distorted images, with the optical data resolving multiple filaments offset asymmetrically from the galaxy.
These are similar to the ``jellyfish'' galaxies in more distant clusters described by Cortese  et al. (2007). In other cases, the features we identify are 
trails of UV emission, in which the optical counterpart takes the form of narrow streams or clumps in the $u$-band image. These objects appear 
more like ``tadpoles'', and include for instance the galaxy with a 60\,kpc H$\alpha$ tail discussed by Yagi et al. (2007). 

\subsection{Comments on individual objects}\label{sec:atlas}

In this section we briefly discuss the characteristics of each galaxy in our sample, and the features which led
to its identification as a gaseous stripping candidate (see also Figure~\ref{fig:images2}).
Where relevant, we summarise any previous literature discussion of each galaxy in the context of gas stripping. 

\begin{figure*}
\includegraphics[angle=0,width=152mm]{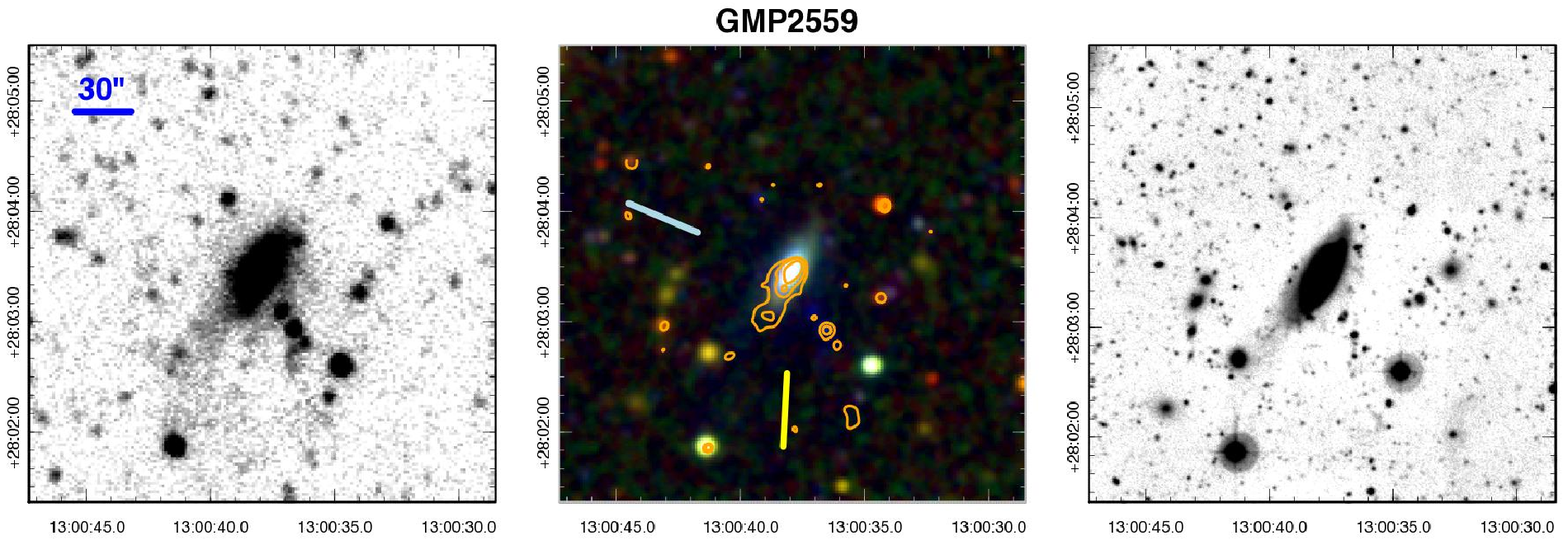}
\includegraphics[angle=0,width=152mm]{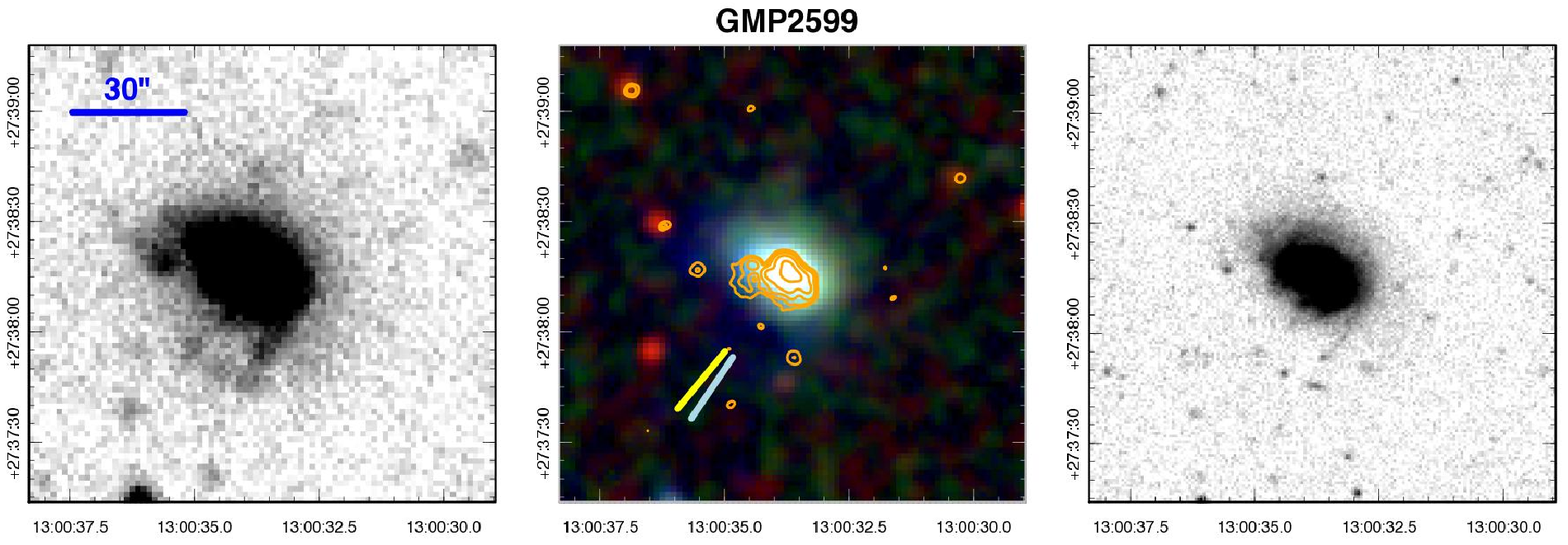}
\includegraphics[angle=0,width=152mm]{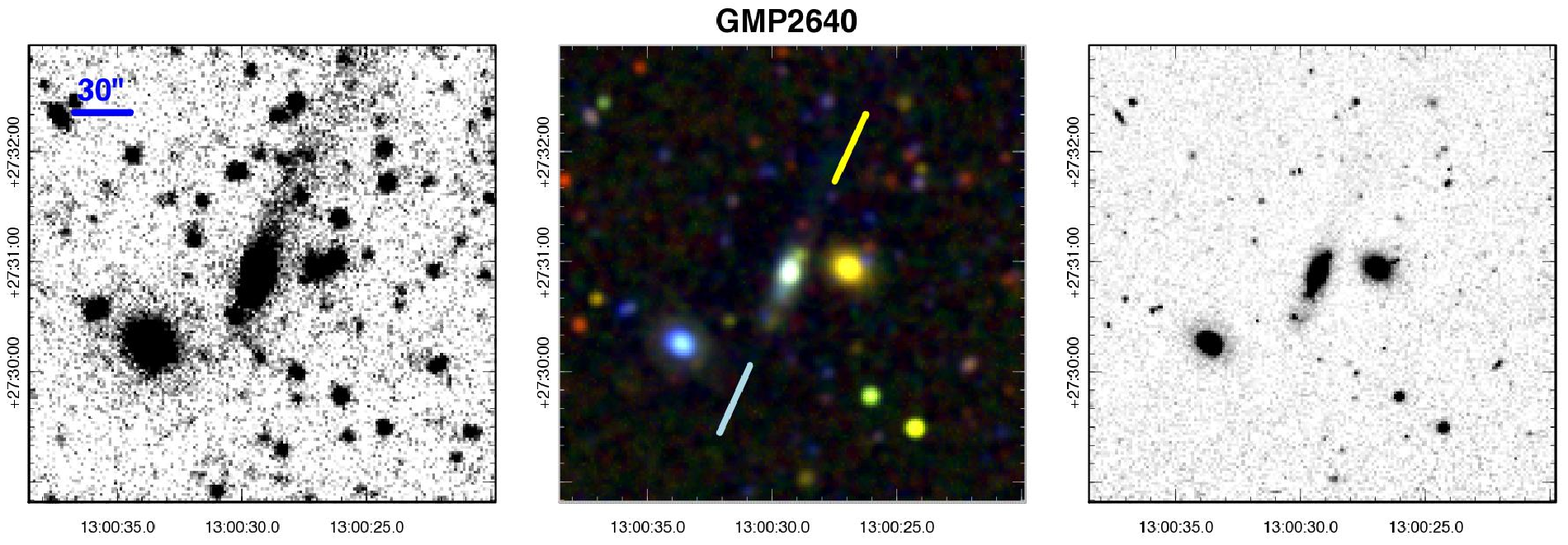}
\includegraphics[angle=0,width=152mm]{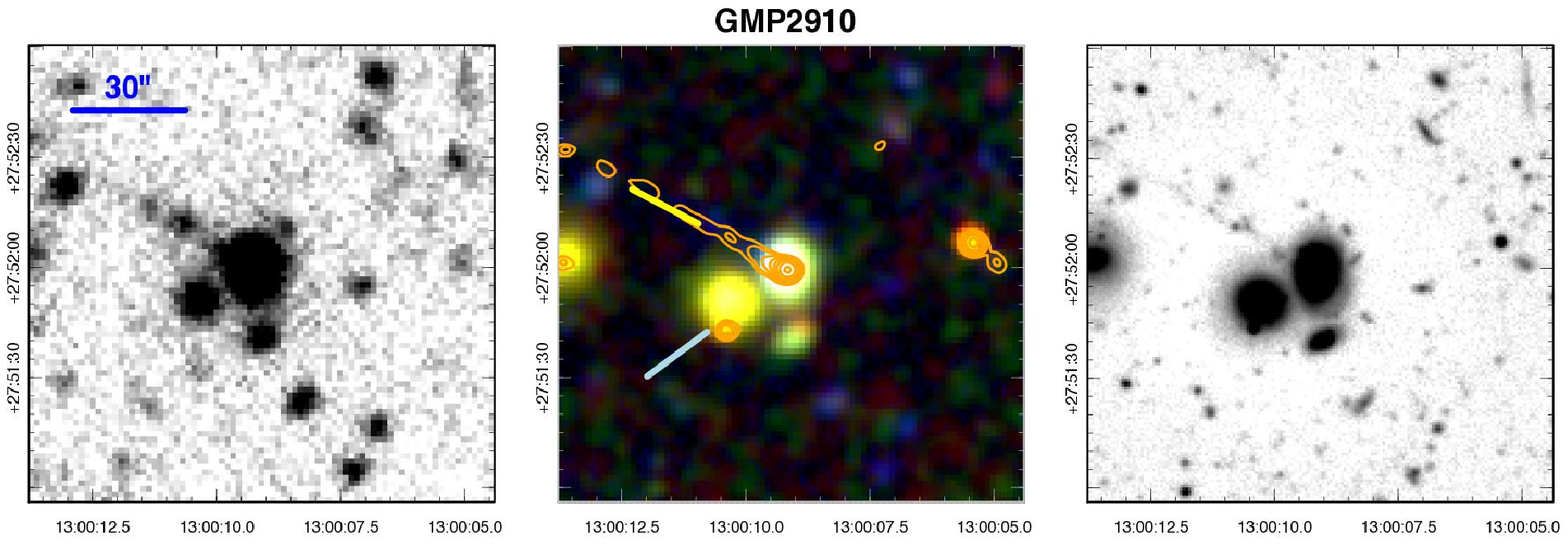}
\caption{Images of the candidate stripping events. For each galaxy, the first panel shows
a greyscale image of the combined {\it GALEX} FUV and NUV bands, while the second panel is 
a {\it GALEX}/MegaCam composite with NUV and FUV combined for the blue channel, $g$ and $u$ combined for the green and 
$i$ for the red.  The yellow line points along the direction we identify as a ``trail'', while the blue line 
points away from the cluster centre. 
H$\alpha$ emission, from INT narrow-band imaging, is overlaid as contours
for selected objects (GMP 2559, GMP 2599, GMP 2910, GMP 3816, GMP 4060, GMP 4471 and GMP 4629). 
The third panel shows the Adami deep MegaCam $u$-band data (except for GMP 5422 where we show  our shallower image).}
\label{fig:images2}
\end{figure*}

\begin{figure*}
\begin{center}
\includegraphics[angle=0,width=152mm]{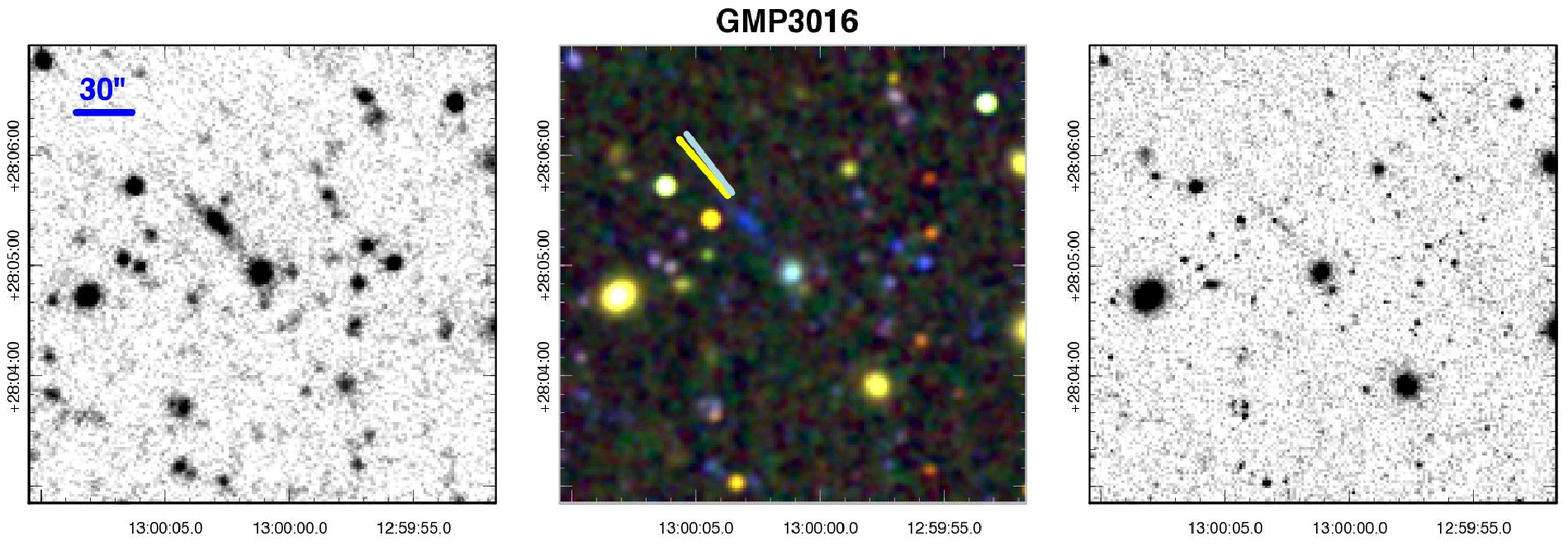}
\includegraphics[angle=0,width=152mm]{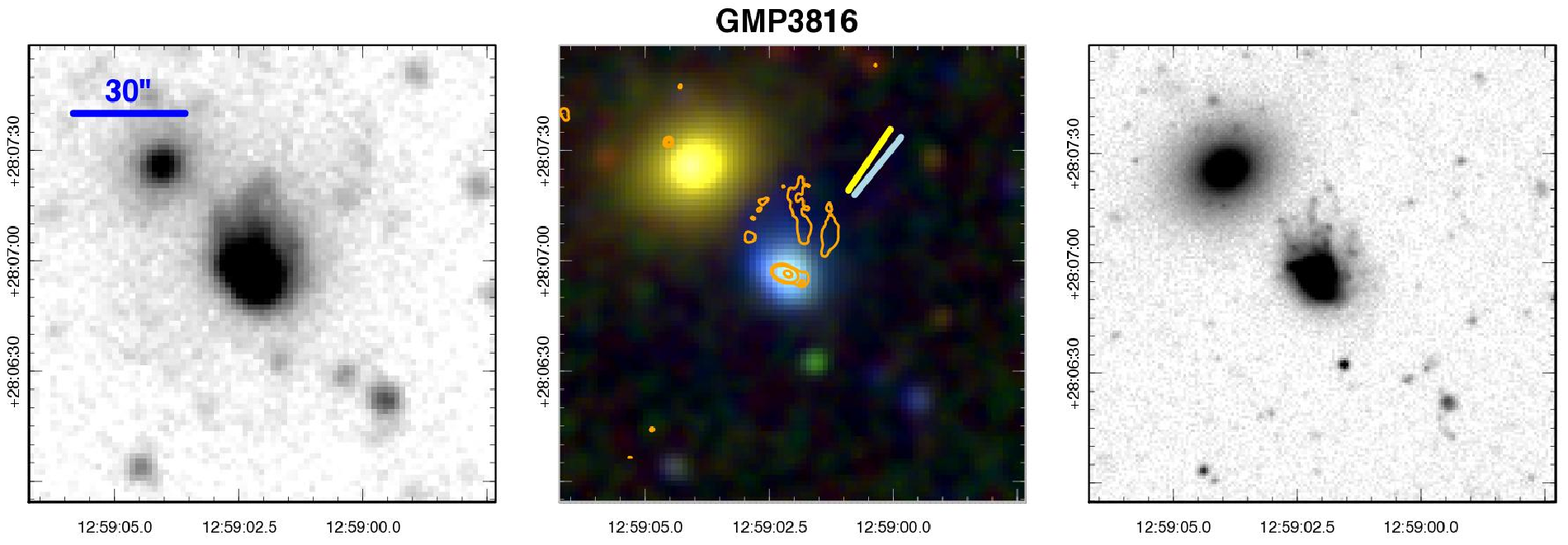}
\includegraphics[angle=0,width=152mm]{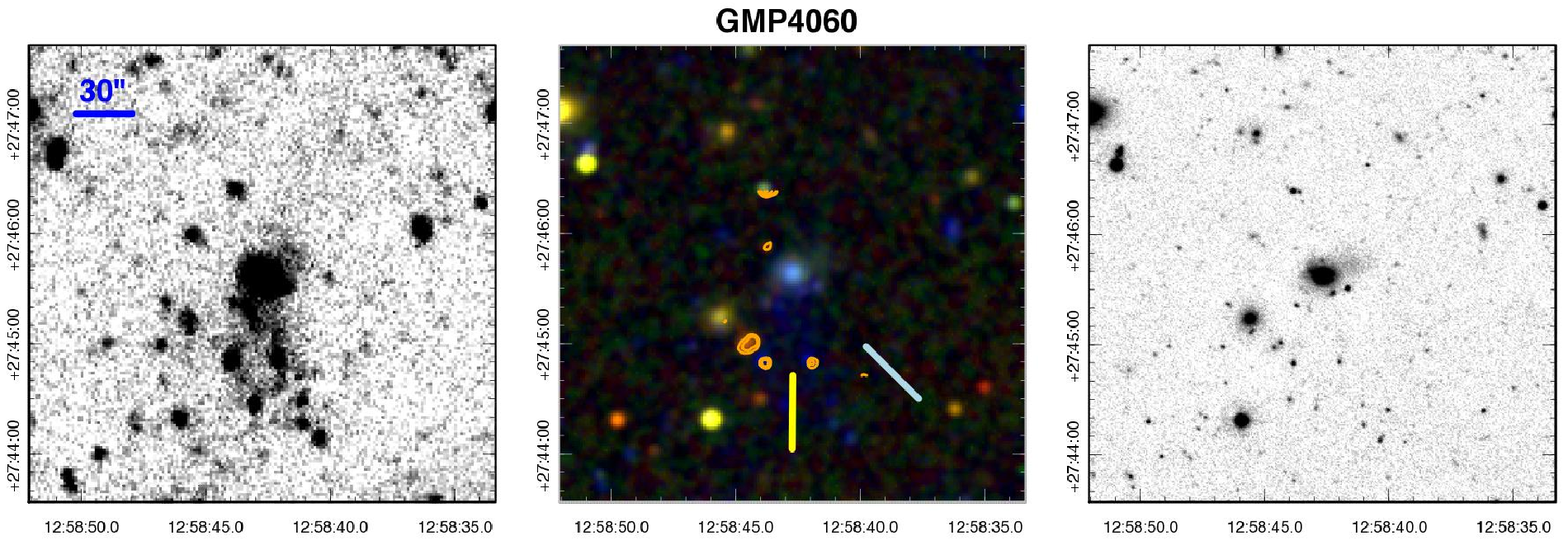}
\includegraphics[angle=0,width=152mm]{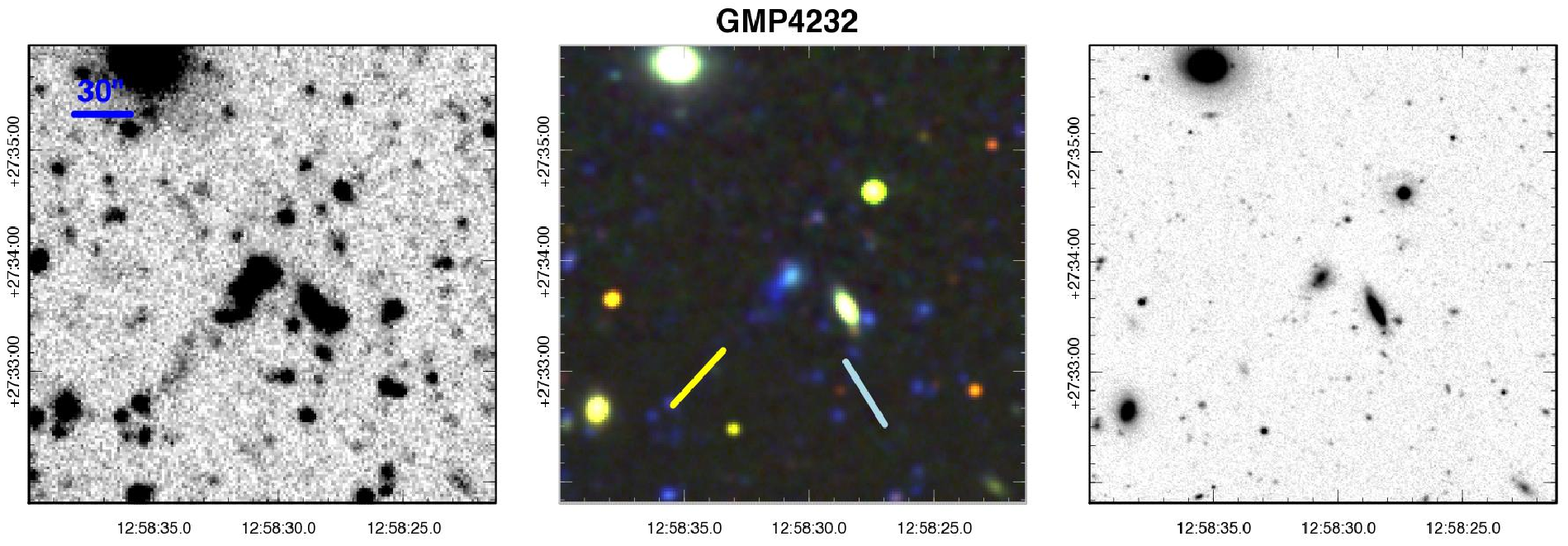}
\end{center}
\contcaption{}
\end{figure*}

\begin{figure*}
\begin{center}
\includegraphics[angle=0,width=152mm]{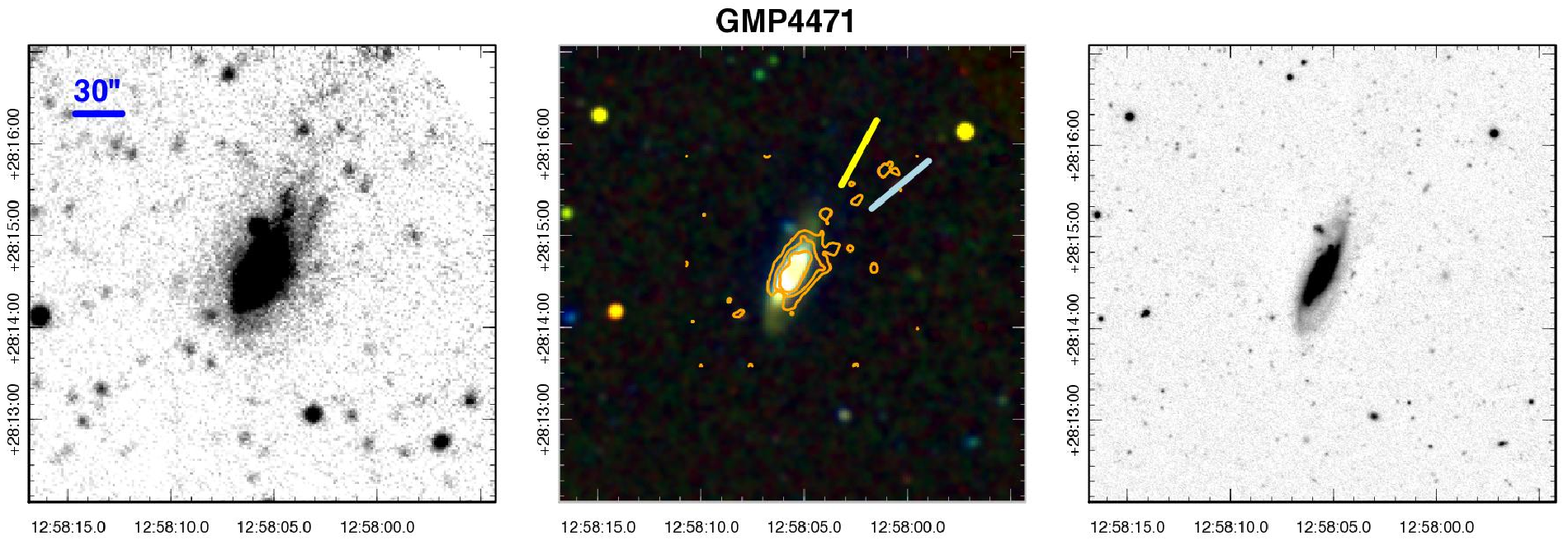}
\includegraphics[angle=0,width=152mm]{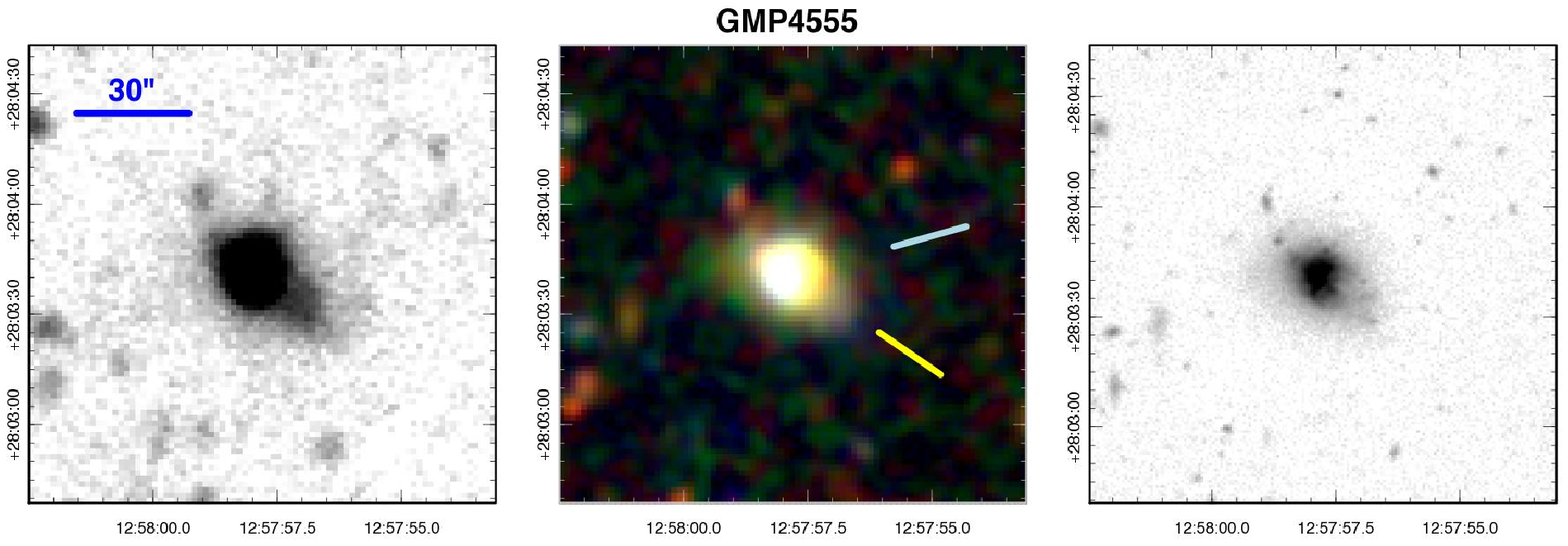}
\includegraphics[angle=0,width=152mm]{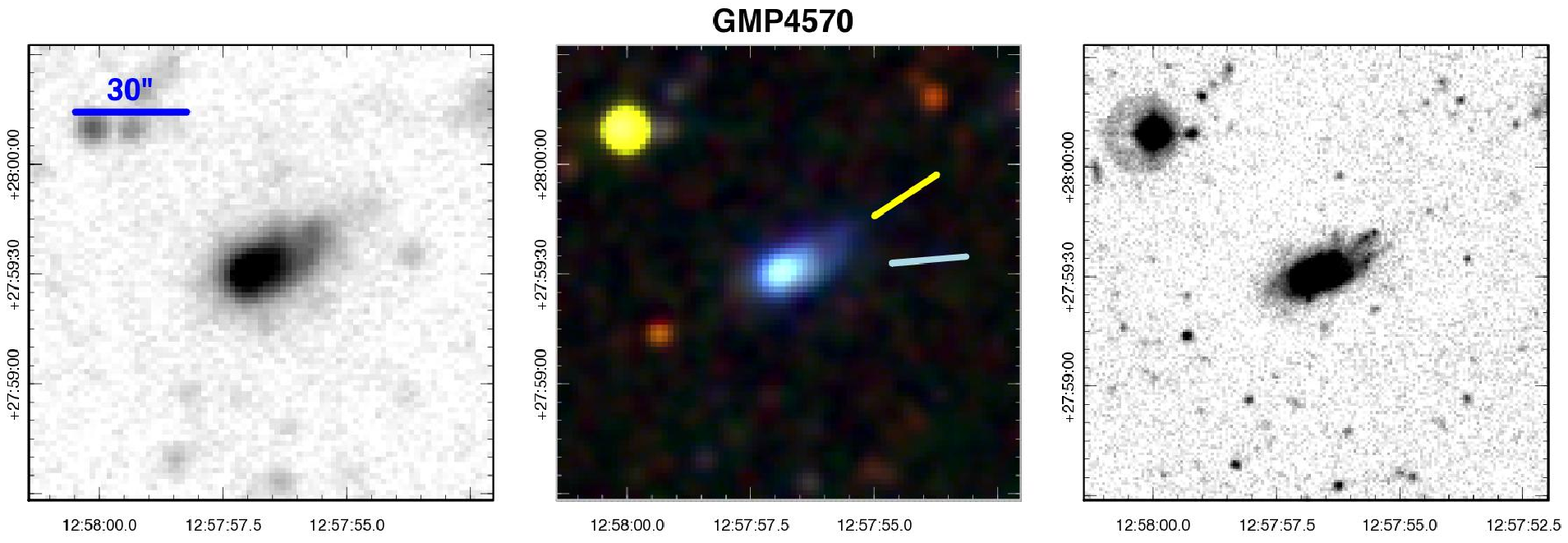}
\includegraphics[angle=0,width=152mm]{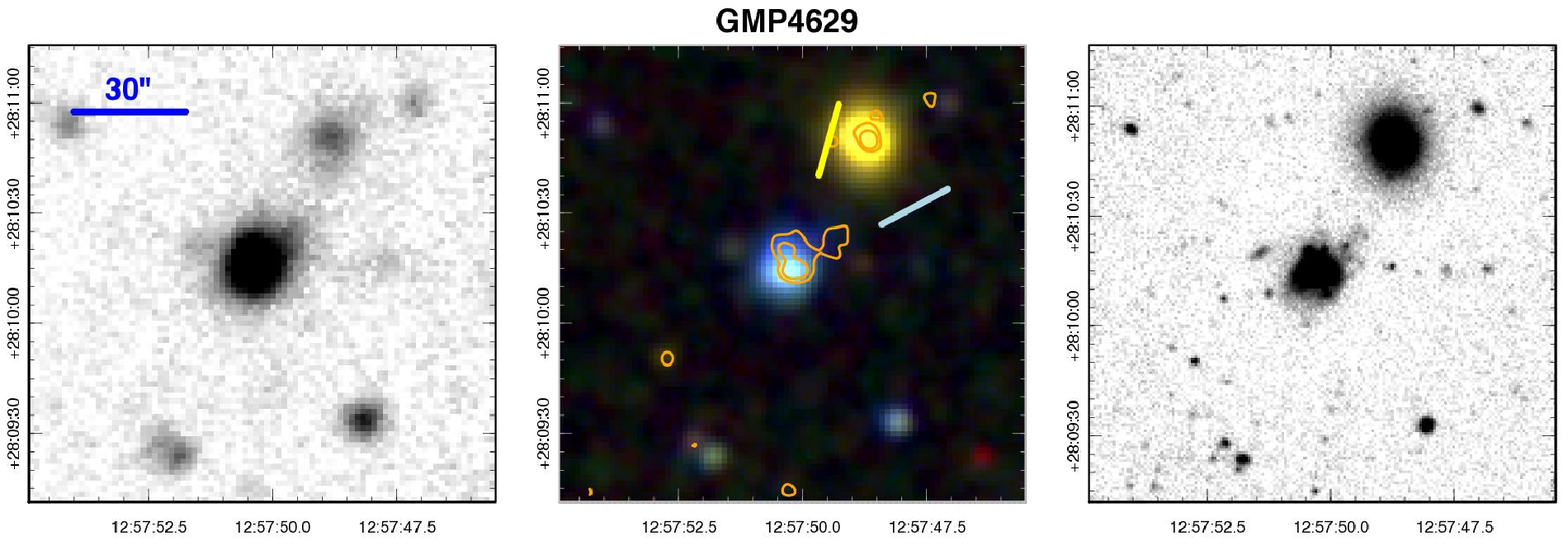}
\end{center}
\contcaption{}
\end{figure*}

\begin{figure*}
\begin{center}
\includegraphics[angle=0,width=152mm]{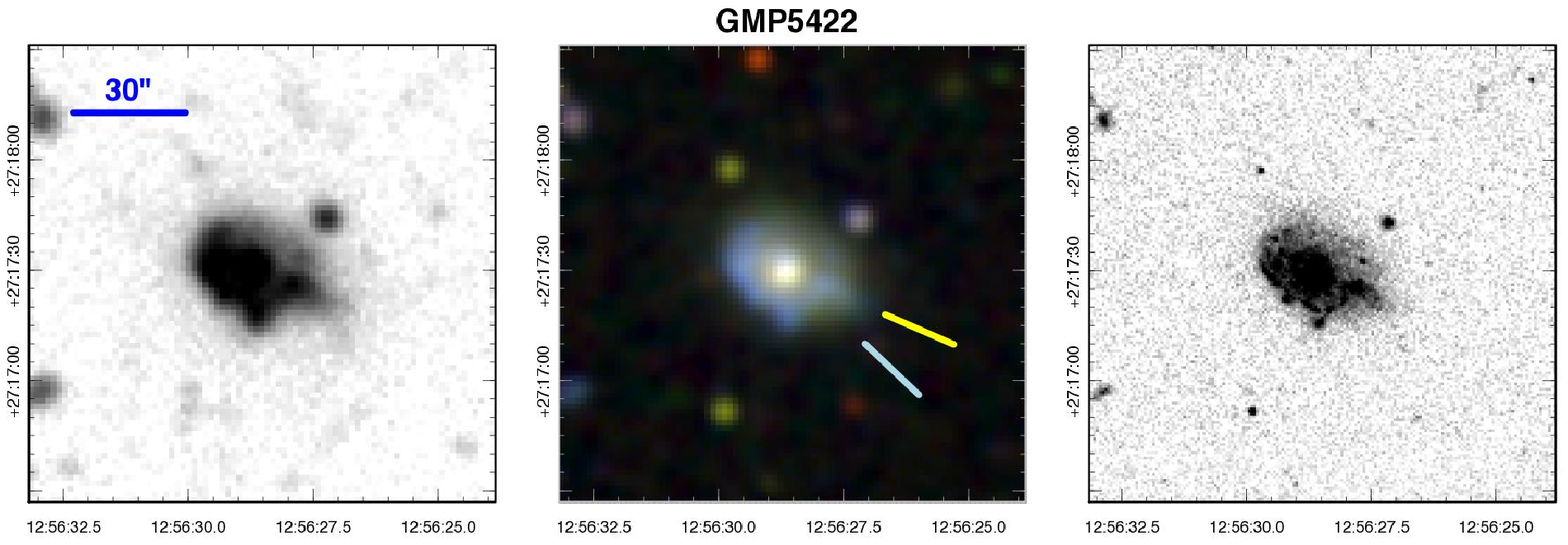}
\end{center}
\contcaption{}
\end{figure*}

\begin{figure*}
\includegraphics[angle=270,width=175mm]{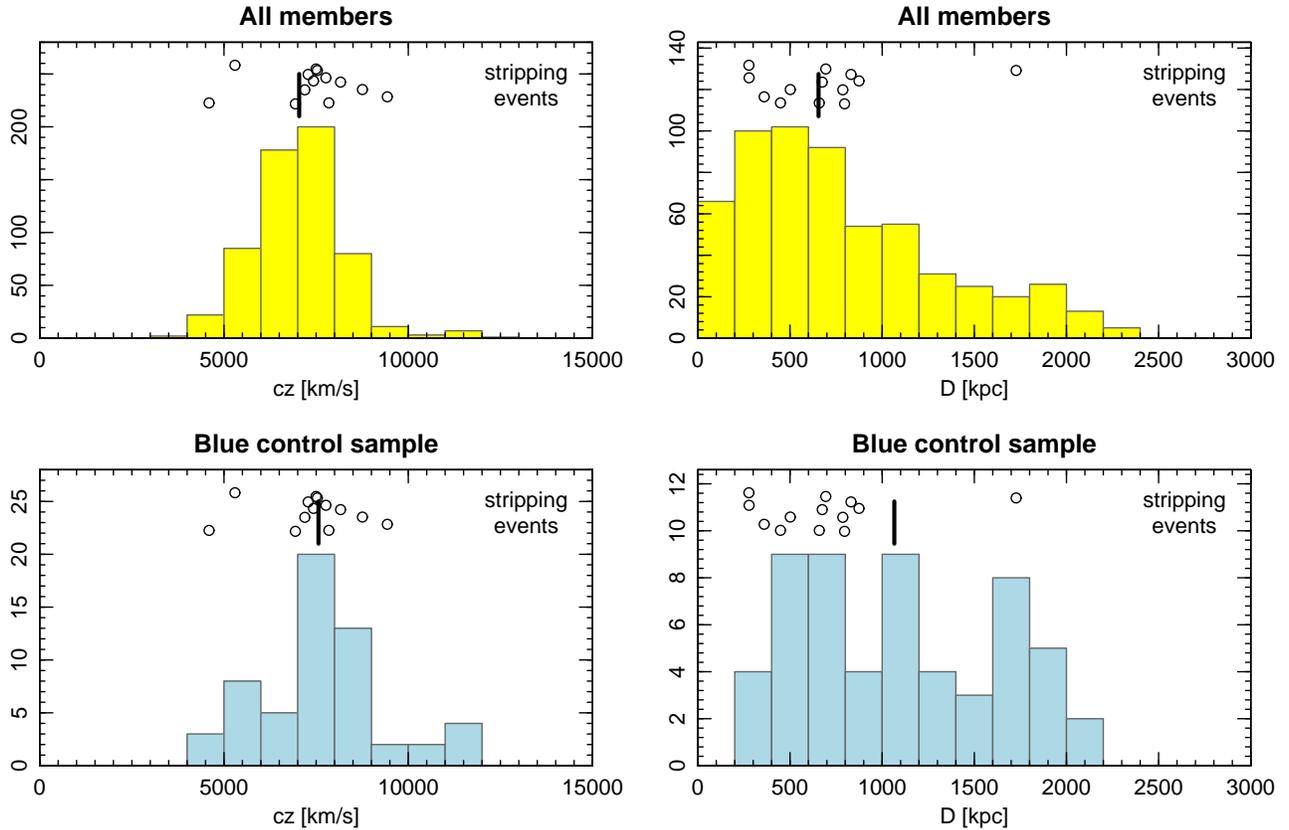}
\caption{The redshift and radial distributions of the galaxies with stripping features, compared to the matched 
Coma member galaxies (upper panels) and to the blue galaxies without ongoing stripping (lower panels).  
The vertical tick above each histogram marks the median of the comparison sample, while the open symbols 
show the stripping galaxies (at arbitrary vertical position).}
\label{fig:histcomp}
\end{figure*}

\subsubsection*{GMP 2559 = IC 4040}

This is a spiral galaxy with an irregular, patchy distribution of dust obscuration. It is one of only two galaxies in our sample that fall
within the footprint of the Coma HST/ACS Treasury Survey\footnote{The custom-reduced ACS imaging from this survey is publicly available
from {http://archive.stsci.edu/prepds/coma}} (Carter et al. 2008, Hammer et al. 2010b). The {\it GALEX} imaging reveals a plume of emission towards the
south-east, and three compact sources to the south-west. At very low surface brightness levels, the region between these two features
appears to be filled by faint diffuse UV emission. 

The UV plume to the south-east extends $\sim$1.8\,arcmin ($\sim$50\,kpc) from the centre of GMP 2559, and can also be seen in the deep 
Adami $u$-band image, though there is no clear counterpart  in our shallower MegaCam data. The trail passes close to a neighbouring
faint red-sequence galaxy GMP 2529, 
which is offset by 775\,\kms\ in radial velocity. Co-located with the UV plume is an H$\alpha$ trail extending at least 1.4\,arcmin (40\,kpc), with 
secondary H$\alpha$ peaks at 12, 23 and 28\,kpc from the nucleus. The ACS imaging appears to show compact, fairly red objects
centrally located within these peaks, but these may simply be chance alignments. 
The galaxy is deficient in HI (Def$_{\rm HI}=0.56$, Gavazzi et al. 2006), and Bravo-Alfaro et al. (2001) note that the HI distribution is offset towards the south-east 
(i.e. the same direction as the H$\alpha$ tail). A displacement in the same sense is seen for both the X-ray and radio continuum emission 
(Finoguenov et al. 2004; Miller, Hornschemeier \& Mobasher 2009).

The three UV-bright objects to the south-west, which have not previously been discussed in the literature, are also seen in the H$\alpha$ 
emission maps, extending 0.8\,arcmin (23\,kpc) from the nucleus. The central object is catalogued as GMP 2572, and has been confirmed 
using spectroscopy from LRIS at Keck (Chiboucas et al., in preparation) to be a strong emission-line source at a velocity of 
7600\,\kms\ (i.e. 250\,\kms\ lower than GMP 2559). Wolf--Rayet features are observed in the spectrum, indicating a very young starburst ($\sim$5\,Myr).
In the Carter et al. (2008) ACS imaging, the peaks of the H$\alpha$ emission to the south-west 
correspond to very blue, irregular systems of compact knots, reminiscent of the ``fireballs" in GMP 4060 (Yoshida et al. 2008)\footnote{The ``fireballs'' in 
GMP 2559 and GMP 4060 are described in greater  detail in Section~\ref{sec:fireballs}.}.
The south-west clumps also appear in archival Spitzer/MIPS data of Bai et al. (2006), as a faint extension of the 24$\mu$m emission from GMP 2559, 
coincident with the UV/optical/H$\alpha$ detections.

\subsubsection*{GMP 2599 = KUG 1259+279A}

This spiral galaxy appears distorted in the {\it GALEX} image, showing broad streaks that point south-east away from the cluster centre. 
Some faint knots are visible in the  MegaCam images.
Miller et al. (2009) note the radio emission is offset from the galaxy and suggestive of ram-pressure stripping, while
Finoguenov et al. (2004) found that the X-ray source is offset to the east of the galaxy. 
Gavazzi et al. (2006) detected the galaxy in HI but indicate that it is strongly gas-deficient (Def$_{\rm HI}=0.76$).
The H$\alpha$ data show an offset in the emission to the east of the galaxy centre. 

\subsubsection*{GMP 2640 = KUG 1258+277}

This appears to be an edge-on S0 or spiral galaxy, which presents a very long trail in the {\it GALEX} imaging, extending 3.5\,arcmin (100\,kpc) 
north towards the cluster centre. In the MegaCam data, the trail is faintly visible and the inner morphology is clearly disturbed. 
The galaxy is an HI non-detection in Bravo-Alfaro et al. (2000), with an upper limit which implies that it is gas deficient  (Def$_{\rm HI}>0.7$).
In the INT data, we do not detect any H$\alpha$ emission, either from the galaxy itself or from the trail.
Close in projection are an elliptical (GMP 2670 at 0.5\,arcmin or 15\,kpc, with 360\,\kms\ radial velocity difference) 
and a spiral (GMP 2601 at 1.2\,arcmin or 35\,kpc, with 1800\,\kms\ radial velocity difference); 
hence this could possibly be a case of tidal stripping by a neighbouring galaxy, rather interaction with the cluster itself.

\subsubsection*{GMP 2910 = MRK 0060 NED01}

This irregular or spiral galaxy has a post-starburst spectrum in the disk region, with the burst age estimated at 250\,Myr 
and ongoing star formation in the nucleus (Caldwell, Rose \& Dendy 1999). 
In our {\it GALEX} image, we observe a narrow tail of length  $\sim$0.5\,arcmin (15\,kpc), extending to the north-east. 
Stripping in this galaxy was first discussed by Yagi et al. (2007), who reported a 60\,kpc H$\alpha$ tail, which is also seen in our
INT H$\alpha$ imaging. This remarkably narrow and straight feature is also clearly seen in the deep Adami $u$-band image, 
and is co-located with the UV trail. The presence of continuum emission suggests that star formation is taking place in the 
stripped material, not merely ionization of a purely gaseous tail as proposed by Yagi et al. 
The galaxy is seen close in projection to an elliptical (GMP 2897, at 0.3\,arcmin or 10\,kpc)
further away is another early-type galaxy (GMP 2852, at 1.0\,arcmin or 30\,kpc), 
but both have large differences in radial velocity
(4700\,\kms\ and 2100\,\kms\ respectively)  and are unlikely to be physically associated with GMP 2910.

\subsubsection*{GMP 3016}

This is among the less convincing cases in the sample. GMP 3016 itself is a faint irregular galaxy. About 0.5\,arcmin (15\,kpc) to the north-east, there is an 
elongated UV source which points towards the galaxy which could be a trail similar to the case of GMP 2910. In the Adami $u$-band imaging, 
the `trail' appears to extend up to 1\,arcmin (30\,kpc), with numerous clumps. A companion is seen in projection to the south-west, 
along the same axis as the `trail'. Its radial velocity is 8314\,\kms, which is 550\,\kms\ larger than that of GMP 3016, so the two galaxies 
are potentially associated, however it is much fainter ($i=18.7$) and unlikely to be causing a major tidal disturbance. 
There is no clear detection of star formation in H$\alpha$.  It is possible that this system could be a chance alignment of unrelated galaxies.

\subsubsection*{GMP 3816 = NGC 4858}

This disturbed barred spiral presents a ``jellyfish'' morphology in the {\it GALEX} image and in the MegaCam data (especially $u$-band).
Several tails and knots, are seen in a broad fan-like distribution, extending 0.5\,arcmin (15\,kpc) to the north-west, away from cluster centre. 
There are strong variations in H$\alpha$-vs-UV flux ratio among the various filaments; in particular the brightest UV structure is not coincident
with the strong central H$\alpha$ feature. The galaxy is deficient in HI (Def$_{\rm HI}>1.1$), based on the upper limit to the gas mass 
published by Gavazzi et al. (2006).
The galaxy is observed close in projection (0.6\,arcmin or 20\,kpc) to a large elliptical, GMP 3792, with 1500\,\kms\ radial velocity difference. It is 
possible that these galaxies are physically interacting, although their relative velocity is quite large compared to the characteristic internal velocities,
and hence strong mutual interactions are unlikely. 

\subsubsection*{GMP 4060}

This galaxy, discussed extensively by Yoshida et al. (2008), has a disturbed optical morphology suggestive of a merger remnant,
and a post-starburst spectrum (Poggianti et al. 2004). There is some tidal stellar debris prominent in the optical images, to the west of the galaxy.
In the {\it GALEX} images, the more spectacular feature is a broad fan of emission extending $\sim$1\,arcmin (30\,kpc) south of the galaxy, with 
several bright knots. Yoshida et al. used deep broadband and H$\alpha$ imaging to reveal a complex network of filaments extending 
up to 2.7\,arcmin 80\,kpc from the galaxy. Embedded within these filaments are bright star-forming knots, which Yoshida et al. refer to as `fireballs',
which in several cases are co-located with the UV emission peaks. The galaxy and part of the filamentary structure fall within the Carter 
et al. (2008) HST/ACS imaging survey, which reveals an extended clumpy substructure within two of the fireballs, very similar to the
knots south-west of GMP 2599 (see Section~\ref{sec:fireballs}). 
GMP 4060 is observed close in projection to a faint elliptical (GMP 4035 at 0.8\,arcmin or 24\,kpc), 
but there is a large radial velocity difference (2100\,\kms), so a tidal interaction is not likely. 

\subsubsection*{GMP 4232}

This is a distorted possibly spiral galaxy with an apparent trail to the south-east in the {\it GALEX} imaging. 
The MegaCam $u$-band image shows a fan of faint filaments/knots 0.5\,arcmin (15\,kpc) 
south-east of the galaxy; beyond this the {\it GALEX} ``trail'' is probably spurious, 
caused by aligned unrelated galaxies. GMP 4232 is close in projection and radial velocity 
to an edge-on S0 (GMP 4255, at 0.6\,arcmin or 20\,kpc, with 290\,\kms\ radial 
velocity difference), so tidal interaction is a possibility.
The INT H$\alpha$ imaging shows one bright knot and some faint extended emission to the SE, coincident with the UV features.

\subsubsection*{GMP 4471 = NGC 4848}

This is the brightest galaxy in the sample, a spiral with a disturbed inner morphology and a faint extension to the north-west 
visible in the MegaCam imaging. In the {\it GALEX} data, we observe trails pointing north-west away from cluster centre, certainly to 
1.4\,arcmin (40\,kpc), and tentatively to twice this distance. (The uncertainty arises because this galaxy is close to the edge of the {\it GALEX} field.)
The INT H$\alpha$ images show an asymmetric distribution of emission in the core, and a stream of emitting knots to the north-west of the
galaxy, some of which are coincident with compact blue sources in the MegaCam imaging. GMP 4471 has been the subject of previous studies:
Gavazzi et al. (1998) showed the H$\alpha$ morphology of the inner region, but no extended structure. Bravo-Alfaro et al. (2001) noted 
an unusual HI distribution peaking 0.3\,arcmin (10 kpc) north-west of the core. It is also deficient in total HI (Def$_{\rm HI}>0.49$, Gavazzi et al. 2006). 
Vollmer et al. (2001) discussed the peculiar morphology in 
CO, HI and H$\alpha$, and proposed this object as a ``post-stripping" galaxy, i.e. one which has already passed through the cluster. 
Finally Finoguenov et al. (2004) noted 
a tail of X-ray emission to the north-west of GMP 4471, making it one of very few galaxies known to exhibit a stripping trail in hot gas. 

\subsubsection*{GMP 4555 = KUG 1255+283}

The {\it GALEX} imaging reveals a UV plume to the west of this disturbed galaxy, oriented away from the cluster centre. 
In the MegaCam $u$-band data, the inner part of the plume is delineated by faint filaments extending to the south-west
The H$\alpha$ emission is also asymmetric and  extended  to the south-west.
Miller et al. (2009) note that the radio emission is offset from the galaxy and suggestive of ram-pressure stripping. 
The X-ray morphology also appears extended to the west in figure 3 of Finoguenov et al. (2004), although they do not specifically discuss this source. 

\subsubsection*{GMP 4570}

This is a distorted spiral galaxy, for which the {\it GALEX} image shows a clumpy UV extension to the west, away from the cluster centre. 
H$\alpha$ is not detected from this object in the INT narrow-band imaging, because its large velocity with respect to the cluster shifts the 
emission line outside of the filter profile. 
The $u$-band MegaCam image shows some trails and knots, coincident with the UV peaks, extending at least $\sim$0.4\,arcmin (12\,kpc) 
from the galaxy centre.

\subsubsection*{GMP 4629}

This distorted spiral shows an asymmetric extension in the {\it GALEX} image. In the MegaCam  data, multiple blue knots are seen to the
north-west of the galaxy, away from cluster centre. The H$\alpha$ imaging shows secondary emission peaks to the north and north-west
of the core. GMP 4629 lies close in projection (0.7\,arcmin or 20\,kpc) to an elliptical galaxy (GMP 4648), 
with 300\,\kms\ radial velocity difference, and hence a tidal interaction is possible in this case.

\subsubsection*{GMP 5422 = IC 3913}

This is perhaps the least convincing example of stripping in our sample. The galaxy is a spiral with a 
distinct opening of the spiral arms to the south-west, and an enhancement of star formation to the north-east. The galaxy is part of  
the NGC 4839 group to the south-west of the cluster core, and lies beyond the extent of our H$\alpha$ observations. 
Previous studies 
in HI (Bravo-Alfaro et al. 2001), in H$\alpha$ (Gavazzi et al. 1998), and in radio continuum (Miller et al. 2009) have not suggested ongoing
stripping in this galaxy.

\subsection{Independent H$\alpha$ identifications}

During revision of this work, Yagi et al. (2010) submitted a paper based on H$\alpha$ imaging of the Coma cluster core from Subaru 
(much deeper than the INT data), in which they {\it independently} select candidate gaseous stripping galaxies based on extended 
H$\alpha$ features. Yagi et al. confirm all six of our {\it GALEX}-selected objects within the region of overlapping observations 
(GMP 2559, GMP 2910, GMP 3016, GMP 3816, GMP 4060 and GMP 4232). It is notable that two of these (GMP 3016 and GMP 4232)
 were among the less secure identifications discussed in the previous section. The Yagi et al. observations hence provide independent 
 support for our sample selection methods in general.

\section{Statistics of gaseous stripping events}\label{sec:stats}

\begin{figure}
\includegraphics[angle=270,width=85mm]{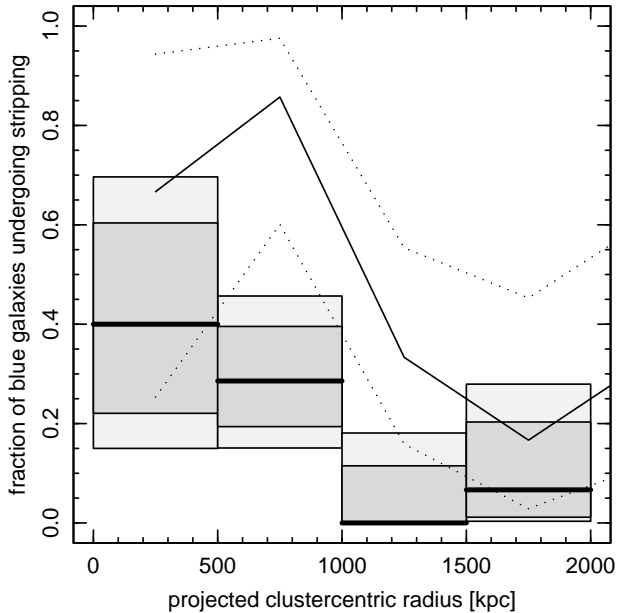}
\caption{The fraction of blue ($NUV-i<4$) galaxies undergoing gaseous stripping, as a function of distance from the cluster centre.
The light and dark intervals represent 90\,per cent and 68\,per cent
confidence intervals, based on binomial statistics. The upper solid line shows the
fraction of bright late-type galaxies from Gavazzi et al. (2006) that are more HI-deficient than 95\,per cent of galaxies beyond 
3\,Mpc. The dotted lines show the 68\,per cent interval on the HI-deficient fraction.}
\label{fig:stripfrac}
\end{figure}

\begin{figure*}
\includegraphics[angle=270,width=150mm]{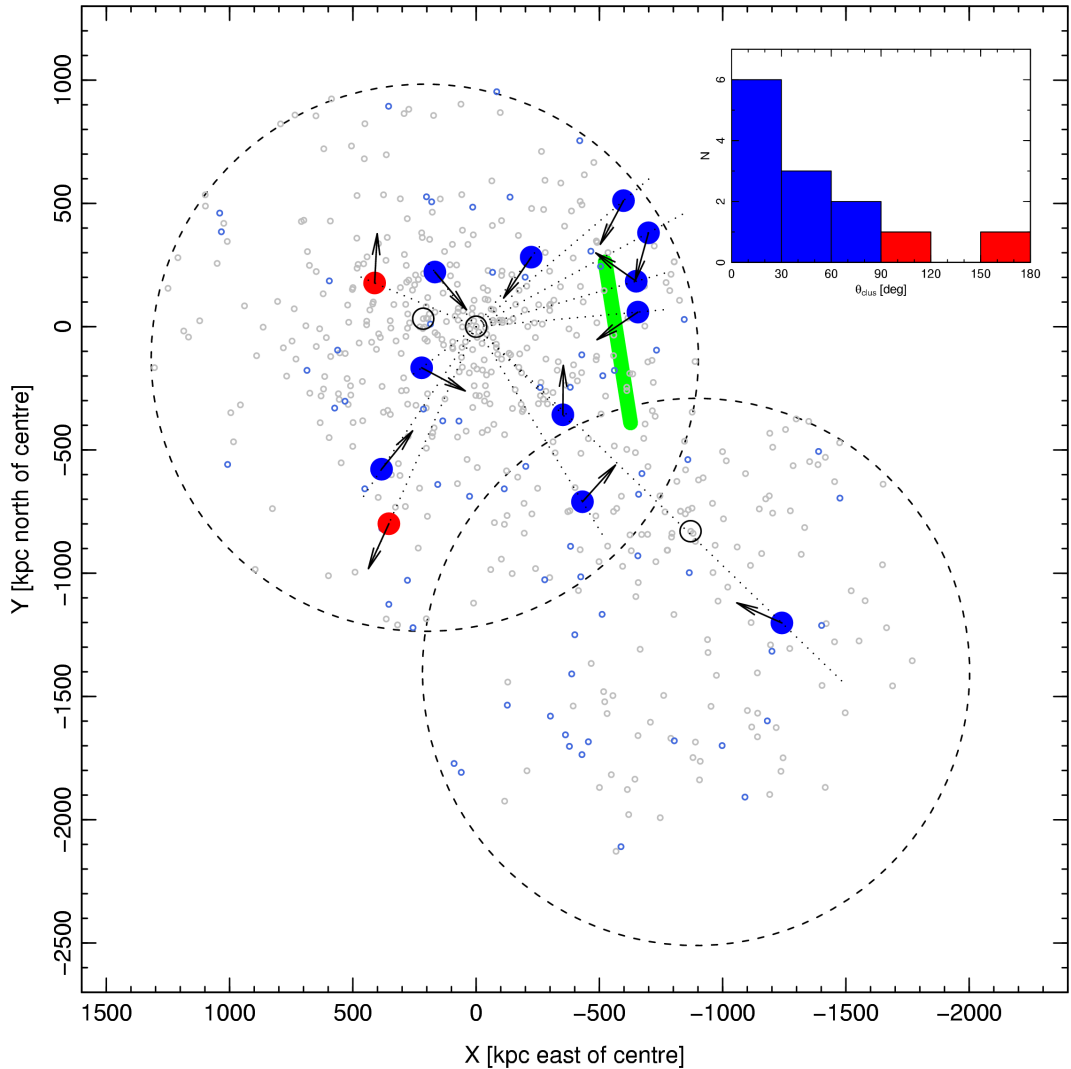}
\caption{The projected distribution and orientation of the galaxies with stripping features. The dashed circles show the extent of the {\it GALEX} imaging, 
while large open circles show the locations of giant ellipticals NGC 4839, NGC 4874 and NGC 4889, for reference. Small circles
show all matched Coma cluster members in our catalogues, highlighting in blue those with $NUV-i<4$. The GSE galaxies are marked
with large filled symbols and vectors denoting the direction of their projected motion, as inferred from their streams or tails. Galaxies
apparently approaching the cluster centre are marked in blue, while those receding from it are plotted in red. Dotted lines indicate the 
cluster-centric direction at the position of each GSE. The thick green line-segment joins the two residual X-ray maxima in the western structure identified
by Neumann et al. (2003).
The inset shows the distribution of  the alignment angle $\theta_{\rm \ \ clus}$ between tails and the cluster-centric vector. We define $\theta_{\rm \ clus}$ to be
near zero for galaxies apparently approaching (i.e. tails pointing {\it away from}) the cluster centre and near 180$^{\circ}$ for galaxies apparently 
receding from (i.e. tails pointing {\it towards}) the cluster centre.}
\label{fig:vecs}
\end{figure*}

The objects identified in Section~\ref{sec:obs} were selected primarily on the basis of asymmetric UV light distribution, indicating a disturbance in the
distribution of recently-formed stars, without corresponding asymmetry in the old stars probed by redder optical bands. 
In general we assume these features to result from interaction with the environment, i.e. either with neighbouring
galaxies, or with the cluster itself, via hydrodynamical or gravitational tidal forces. 
These forces act on {\it gas} in the parent galaxies, not only stripping gas from the galaxy disks but also triggering the formation of 
new stars which become visible in the UV. In the hydrodynamical case, only the gas is directly affected, but the newly-formed
stars provide the observed evidence. In the case of tidal interactions, existing stars are affected as well as the gas, 
but the formation of new stars, fundamentally a gaseous process, causes the asymmetry to be especially prominent in the UV. 

We therefore describe the UV trails discussed above as being due to ``gaseous stripping events'' (GSEs hereafter): the production of UV 
asymmetries occurs through stripping of gas, star formation within that gas, and subsequent UV emission from the new stars. 
The term GSE, as we use it here, does {\it not} distinguish between various possible physical mechanisms: tidal interaction with 
neighbours or with the cluster potential, ram-pressure stripping, viscous stripping etc (see Boselli \& Gavazzi 2006 for detailed discussion
of these processes). In what follows, however, we muster support for a picture in which ram-pressure stripping is a major (though not necessarily the only) 
process responsible for the objects we observe. 

Instead of considering the details of {\it individual} galaxies, our approach is to use the extensive data for Coma to analyse the ensemble properties
of the {\it sample}. In this section, we examine the incidence and distribution of the GSE galaxies, in comparison to the cluster galaxy
population at large, which is dominated by the red sequence members.
We also compare the GSEs to a blue control sample comprised of the 57 galaxies with $NUV-i < 4$ which
are {\it not} identified as GSEs (after excluding the ten galaxies judged to have unreliable colours during the visual inspection). 
Implicitly we assume that these star-forming galaxies are the ``parent'' population from which the GSEs were triggered, 
and therefore statistical differences between the GSEs and control sample indicate the conditions responsible for the stripping events.

The left-hand panels of Figure~\ref{fig:histcomp} show the redshift distribution of the GSE galaxies, compared
to the other Coma cluster members. The GSE galaxies on average have slightly higher radial velocities 
than the full sample of matched members, but given the large velocity dispersion ($\sim$1000\,\kms) in Coma, 
the offset is not significant  ($316\pm370$\,\kms). 
The sample of blue members not undergoing gaseous stripping is offset in the same sense, 
relative to the galaxy population at large (by $560\pm250$\,\kms). Thus the GSEs are consistent with having been drawn from
the same velocity distribution as either the full sample or the blue control sample.

The right-hand panels of Figure~\ref{fig:histcomp} present equivalent comparisons for the radial distribution, adopting a cluster
centre coincident with the giant elliptical galaxy NGC 4874. The GSE galaxies are markedly more concentrated towards the cluster 
core than are the blue control galaxies, with 12/13 (92\,per cent) lying within 1\,Mpc, compared to 26/57 (47\,per cent) of the control sample. 
A Kolmogorov--Smirnov (KS) test yields a $0.4$\,per cent probability that the GSE galaxies were drawn from the same radial distribution as the 
control sample galaxies. The GSEs are, in fact, consistent with being drawn from the much more concentrated distribution followed by all
matched cluster members, which is dominated by passive galaxies.
Note also that the outermost GSE galaxy is GMP 5422 which is the least certain case of stripping identified here. Removing it from the 
sample would {\it strengthen} the discrepancy between GSE and control-sample galaxies in terms of the radial distribution.

Next we consider the fraction of all star-forming ($NUV-i<4$) galaxies which are currently undergoing gaseous stripping.
For the surveyed region as a whole, the GSE galaxies form a fraction $N_{\rm GSE}/N_{\rm blue}=13/70=0.19^{+0.06}_{-0.05}$. 
However, as shown above, the incidence of stripping is concentrated towards the cluster core. Within a radius of
1\,Mpc from the cluster centre, the fraction becomes $N_{\rm GSE, 1Mpc}/N_{\rm blue, 1Mpc}=12/38=0.32\pm0.07$. 
These fractions refer to the stripping features detectable in our data; future deeper imaging might reveal even an larger incidence
of low-level stripping activity. 
Figure~\ref{fig:stripfrac} shows graphically the variation in stripping fraction with radius. Clearly we have insufficient numbers to 
draw any secure conclusions regarding the form of the decline, e.g. whether there is really a sharp cut-off somewhere around 1\,Mpc, 
which could indicate a threshold ICM density for stripping.

We have visually estimated the position angles of the streams/trails/tails, 
and calculated the angle relative to the cluster-centric vector for each galaxy, again adopting 
the position of NGC 4874 for the cluster centre. 
In some galaxies the position angle of the debris is unclear, and a the adopted angle is open to debate.
For the case of GMP 2559, we interpret both the south-east trail and the south-west clumps as the brightest parts of 
a broad fan of stripped material, and hence assign a position angle which is close to south. For GMP 4060, we ignore the tidal 
stellar debris to the west, and assign a position angle close to south, to describe the star-forming knots and filaments. 
In each case, the position angles are shown in the images
of Figure~\ref{fig:images2}. Figure~\ref{fig:vecs} shows the location and orientation
of all the GSEs within the cluster, and summarises the distribution of alignment angles.
Among the thirteen objects selected here, six show streams extending {\it away from} the cluster centre 
(within 30$^\circ$ of the cluster-centric vector), compatible with stripping occurring on passage towards the cluster centre. 
Only one galaxy (GMP 2640) has a stream pointing directly {\it towards} the cluster centre, which instead implies that the galaxy has
already passed through the cluster core. 
In the remaining six objects, the asymmetries are at large angle to the cluster-centric axis, but favour 
galaxies approaching, rather than receding from the cluster core.
The distribution of cluster-centric angle is incompatible with a uniform distribution
at the 99\,per cent level, according to a KS test.

\section{Discussion}\label{sec:disc}

In this section, we discuss our results in the context of other observations of stripping in dense environments. In particular, we
compare with other cases of UV/H$\alpha$ trails in other rich clusters (Cortese et al. 2007; Sun et al. 2010),
and with observations of HI tails in Virgo (Chung et al. 2007, 2009). We also make reference to simulations, especially
those of Kapferer et al. (2009), to interpret the various observed structures.

\begin{figure}
\includegraphics[angle=270,width=85mm]{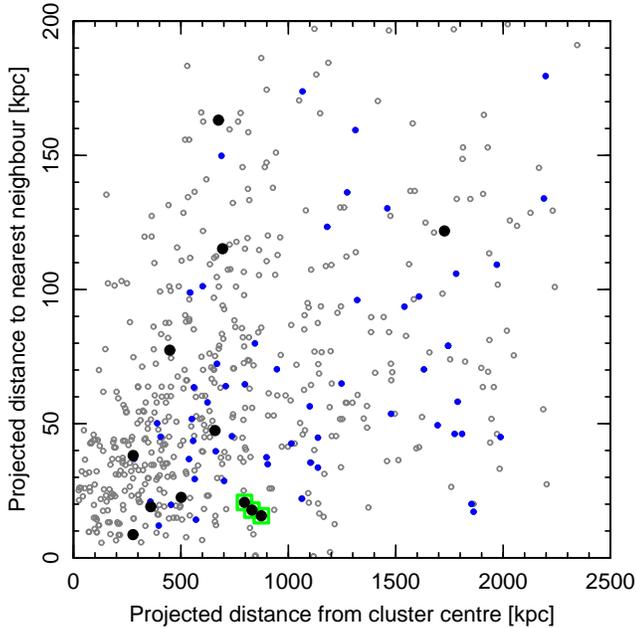}
\caption{Projected distance to the nearest bright neighbour for cluster members, as a function of distance from the cluster centre. To identify 
potential cases of tidal interaction, we select only the neighbours that are no more than 0.5\,mag fainter (in $i$) than the target galaxy.
The overall broad correlation arises because the density of cluster members is highest near the cluster centre. 
The black points show the GSE galaxies, the blue points are the control sample of blue cluster members, and the grey points are the red cluster members. 
The points highlighted in green are GSE galaxies with bright neighbours close in projection, and at similar radial velocity (offset $<500$\,\kms).
}
\label{fig:neighdist}
\end{figure}

\subsection{Ram-pressure stripping on first passage into cluster}

The key results from Section~\ref{sec:stats} are that blue galaxies with tails and other stripping features are preferentially located closer to 
the cluster core than blue galaxies without such features, and that the stripped tails predominantly point away from the cluster centre. 
Both results suggest that the stripping events are in general triggered by interaction with the cluster itself; while 
ram-pressure stripping by the intra-cluster gas is probably the key process involved, other mechanisms such as tidal interactions
may contribute to the features observed. 

As noted in Section~\ref{sec:atlas}, it is possible that at least {\it some} of the GSE galaxies are interacting with 
neighbouring galaxies. 
Such interactions are not simply an alternative to ram-pressure stripping, but may act in combination with it. 
Simulation work (e.g. Vollmer 2003; Kapferer et al. 2008) has demonstrated that gravitational interactions between galaxies can act to 
enhance the efficacy of ram-pressure stripping, by pulling gas to larger radius where it can be more easily removed by the intra-cluster
wind. 
In Figure~\ref{fig:neighdist}, we show the projected distance to the nearest neighbour for galaxies in the matched member 
sample. The neighbours are drawn from the same sample, i.e. confirmed cluster members with $M_i<-17$, and further restricted to be no
more than 0.5\,mag fainter (in the $i$ band) than the ``target'' galaxy, 
since tidal stripping is most efficient for companions of comparable mass. The figure demonstrates that {\it most} of the GSE candidates do 
not have closer neighbours than non-GSE galaxies at similar distance from the cluster centre.
However, in three cases (GMP 2640, GMP 4232 and GMP 4629) the nearest neighbour is very close and also 
has similar radial velocity to the target galaxy, as required for tidal interactions. 
(The other three galaxies with close projected neighbours  have velocity differences $>$1500\,\kms, and are therefore unlikely to be physically interacting.) 
GMP 2640 is the single clear exception to the the tendency for tails to be aligned away from the cluster centre, so it is tempting to invoke tidal 
stripping of stars and/or gas, rather than (or in combination with) ram-pressure stripping to account for this galaxy. Finally GMP 4060 is worthy 
of note here; it is probably a merger remnant, and thus a case in which ram-pressure and tidal interaction
may have acted together to form its spectacular system of knots and filaments.
We conclude from this test that tidal interaction with neighbouring galaxies is not the mechanism 
responsible, {\it in general} for the stripping events identified in this paper, but may play a role in some specific cases. 

An alternative possibility which is more difficult to distinguish is that the stripping may be due to tidal interactions, not with neighbouring galaxies, but with the 
cluster potential itself. Such effects would  produce disturbances oriented preferentially along the cluster-centric vector, as in the case of ram-pressure stripping. 
There is qualitative evidence against {\it purely} tidal interaction, in that the old stellar material, as traced by the $i$-band luminosity distribution, 
appears fairly undisturbed in most of the GSE galaxies, but this is hard to quantify. Moreover, as shown by Smith et al. (2010) from a {\it GALEX}
atlas of interacting field galaxies, the triggering of star formation from gas in tidally-stripped material can lead to tails that are accentuated in the UV even in
the absence of ram-pressure stripping. Ultimately, the head--tail morphologies of many of the GSE galaxies provide the strongest, though not conclusive, evidence that the stripping events are indeed caused by ram pressure, rather than tidal interaction.

Interpreting the observed trails as stripped material {\it behind} the galaxies that have suffered gaseous stripping, 
we infer that the stripping events are occurring mainly on approach to the cluster centre. Furthermore, the absence of GSEs moving 
away from the cluster centre indicates strongly that the stripping event is seen on {\it first} 
approach to the cluster centre, and generally completed on a time-scale short compared to the crossing time.
Combined with the typical cluster-centric radius of 700\,kpc, and a velocity dispersion of 1500\,\kms, 
this yields a time-scale for the stripping event of $T_{\rm GSE}\la500$\,Myr. 

This crossing-time argument is supported by analysing more realistic infall trajectories extracted from the Millenium Simulation (Springel et al. 2005). 
We selected all galaxies with stellar mass greater than $10^{9}\,M_\odot$ in the model of Font et al. (2008), that are members of the 
four most massive clusters in the simulation (halo masses $\sim10^{15}M_\odot$). 
Ignoring the star-formation histories assigned by the semi-analytic machinery, we instead construct a simple model 
which assumes each galaxy enters the cluster forming stars (i.e. is ``blue'') and is then stripped of gas when it first comes within  
a threshold distance of 1\,Mpc from the centre of the cluster. The galaxy is assumed to be observable as an ongoing stripping
event (i.e. a GSE) for a period $T_{\rm GSE}$ after crossing this threshold radius, and to remain ``blue'' for a longer interval $T_{\rm blue}=1.5$\,Gyr after 
stripping begins, while young stars are still present. Identifying the stage reached in this process for all galaxies at $z=0$, the predicted ratio of 
GSEs to all blue galaxies can be compared to our observed GSE fraction as a function of projected radius from the cluster centre. 
To match the assumptions used in interpreting the observations, we use the angle between the projected velocity vector
and the direction to the cluster centre to determine whether the simulated GSEs are apparently approaching or receding from the cluster centre.

\begin{figure*}
\includegraphics[angle=270,width=180mm]{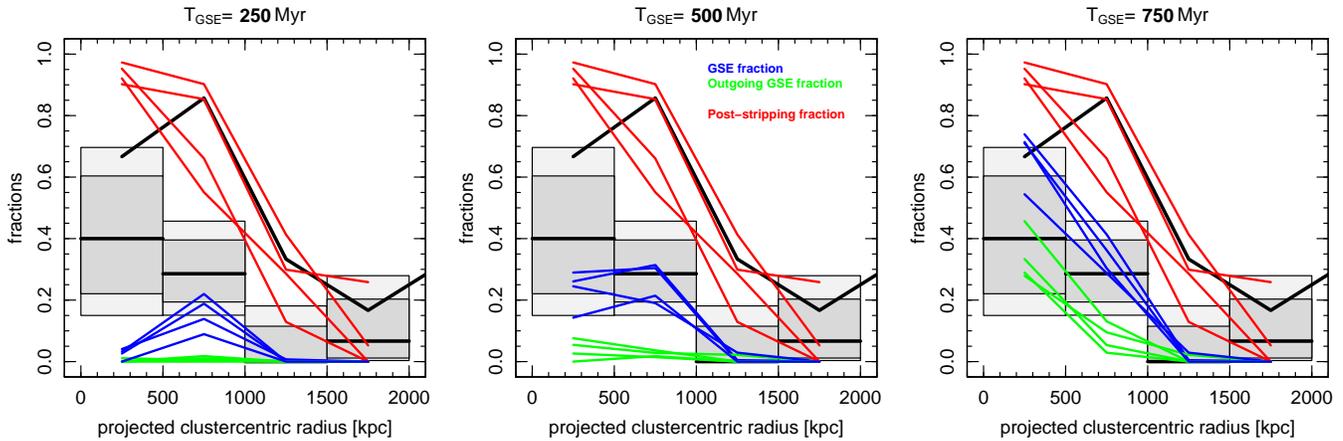}
\vskip 2mm
\caption{A comparison between the observed stripping fraction and predictions from a simplistic model based on galaxy orbits in the four richest
clusters of the Millenium Simulation. The model assumes that galaxies enter as ``blue'' objects, become visible as ongoing stripping events for a 
duration $T_{\rm \ \ GSE}$ after first passing within 1\,Mpc of the cluster centre, and turn from ``blue'' to ``red'' 1.5\,Gyr after crossing this threshold. 
The blue lines (one for each cluster) show the fraction of all blue galaxies that are undergoing stripping at $z=0$, for comparison with the
observed GSE fractions (grey boxes as in Figure~\ref{fig:stripfrac}).  The green lines show the ongoing stripping events that are apparently
moving away from the cluster centre (only $<$10\,per cent of all GSEs in the observed sample). The red lines show the fraction of all
``blue'' galaxies that have passed through the stripping threshold radius, which may be compared to the observed HI-deficient fraction (solid
line as in Figure~\ref{fig:stripfrac}, the error bounds are not shown here, for clarity). Only for stripping time-scales of $T_{\rm \ GSE}\approx500$\,Mpc can we reproduce the fraction of GSEs without generating many more ``outgoing'' GSEs than are observed. 
}
\label{fig:millengse}
\end{figure*}

The results of this analysis are shown in Figure~\ref{fig:millengse}, for three values of the stripping time-scale $T_{\rm GSE}$. As expected, 
the total fraction of model GSEs increases with with increasing $T_{\rm GSE}$. The fraction of ``outgoing'' GSEs is small 
until $T_{\rm GSE}$ becomes comparable to the time taken by an typical galaxy to travel from the 1\,Mpc threshold radius to the peri-centre of its orbit. 
To match the observed total GSE fraction of $\sim$30\,per cent, we find that 
a stripping duration longer than $\sim$300\,Myr is required.  However, the duration must be also be shorter than $\sim$700\,Myr, 
to avoid predicting too many outgoing GSEs. An intermediate time-scale of $T_{\rm GSE}\approx500$\,Myr, yields acceptable
agreement, with $\sim$25\,per cent of blue galaxies undergoing stripping, and $\sim$80\,per cent of GSEs approaching the cluster centre,
although these averages mask substantial cluster-to-cluster variation.
The threshold radius at which the GSE phase is assumed to begin (held at 1\,Mpc in all panels of Figure~\ref{fig:millengse})
is constrained by the radial distribution of GSEs and by their total fraction. The threshold radius must be larger than $\sim$750\,kpc to 
match the GSE fraction in the second bin. The threshold radius must also be smaller than $\sim1.5$\,Mpc; otherwise either the fraction 
of GSEs falls too low at low radius (for short GSE time-scale) or else the overall fraction of GSE is much too large (for longer GSE time-scales). 

In summary, this fairly crude modelling suggests that the fractions of objects undergoing stripping are consistent with what may
be expected from galaxy accretion paths within clusters, and yields approximate constraints on the threshold radius 
and duration of the observed stripping events.

\subsection{Typical density for gas stripping}

In the UV, we observe gaseous stripping occurring predominantly within a projected radius of 1\,Mpc. For Coma this corresponds,
using the beta-model fit of Briel et al. (1992), to a hot gas density of $\rho\approx10^{-27}$\,g\,cm$^{-3}$. 

For comparison, the ``jellyfish'' galaxies identified by Cortese et al. (2007) in clusters at $z=0.2$ are located at projected distances of $\sim$300\,kpc
from their cluster centres, corresponding to $\rho\approx10^{-25}$\,g\,cm$^{-3}$ using the beta-model parameters tabulated in their paper.
ESO137--001 in Abell 3627 is projected $\sim$200\,kpc from the cluster centre, where the ambient density is 
$\rho\approx10^{-27}$\,g\,cm$^{-3}$ (Sun et al. 2010). 
The galaxy in Abell 2125 discussed by Owen et al. (2006) is also close to the core, at a projected distance of $\sim$100\,kpc. 
Since these are all very rich clusters, these projected radii are directly comparable with Coma, and we may conclude that our GSE galaxies 
typically lie at larger radii (both absolute and relative to viral radius), by a factor of $\sim$3 than these previous examples. 
Of course, previous detection of GSEs at large radii will have been limited by  the requirement for wide fields of view 
(especially for nearby clusters) or HST observations (for distant clusters), introducing a bias towards
the most central objects. Moreover, the density required for effective stripping will likely depend on galaxy mass, so that 
the giant galaxies studied in previous work will undergo stripping only in the centres of clusters, while the much fainter 
objects identified here (typically $M_i^\star+1.5$) are susceptible to gas loss at larger radii. 

Our results can also be compared to the incidence of neutral gas tails identified behind spirals in the Virgo cluster by 
Chung et al. (2007, 2009). From a survey sampling cluster-centric radii 0.3--3.3\,Mpc, they identify seven such galaxies, all of 
which lie at intermediate radii, 0.6--1.0\,Mpc. The HI tails are all pointing away from the centre of Virgo. 
The radii are similar in absolute terms to those of our GSEs in Coma, but relative to the corresponding cluster virial radii, 
the HI tails in Virgo are located further out in the cluster, by a factor of $\sim$4, and at much lower external gas density, 
$\rho\approx2\times10^{-28}$\,g\,cm$^{-3}$.
A simple interpretation of this comparison would be that removal of neutral gas and production of HI tails can occur at a lower
ICM density, while higher densities are required to trigger significant star formation in the stripped material. This is supported
by the simulations of Kapferer et al. (2009), whose figures 6 and 7 show the fraction of all newly-formed stars that are located in 
the wake. For ambient densities $\rho=10^{-28}$\,g\,cm$^{-3}$, and relative velocity 1000\,\kms, this fraction is $\sim$10\,per cent, 
while for $\rho>10^{-27}$\,g\,cm$^{-3}$, a majority of the triggered star formation occurs in the wake.

Finally, we consider whether {\it localised} enhancements of the ICM might be responsible for the stripping events we observe. 
Analysing the XMM mosaic observation of Coma, Neumann et al. (2003) showed the presence of an extended enhancement in 
emission (relative to a symmetric model) in the west of the cluster, at projected radii 400--900\,kpc, with two maxima in the X-ray residual map. 
Four of our GSE galaxies lie just 100--200\,kpc beyond the north-west maximum. Although we cannot yet quantify this association, it is
at least suggestive that some stripping events could be triggered by encounters with local density enhancements in the ICM. 
This possibility is reminiscent of the claim by Poggianti et al. (2004) that young post-starburst galaxies in Coma 
are related to local features in the ICM, indeed including the same Neumann et al. western structure. 
The most prominent X-ray substructure in the region we have studied is the group centred on NGC 4839 to the south-west of the cluster core. 
Only one GSE galaxy, GMP 5422, is identified in this region, and as noted in  Section~\ref{sec:atlas}, this is among the least certain cases. 
Inspecting the XMM mosaic, the local projected X-ray surface brightness close to GMP 5422 is comparable to that at $\sim$1\,Mpc  elsewhere 
in the cluster. If  GMP 5422 were to be confirmed as a GSE, it would provide further evidence for local ICM structures as the 
driver for gaseous stripping in clusters.

\subsection{Intra-cluster star formation}\label{sec:fireballs}

We now turn to consider the fate of the stars which form in the stripped material. 
If the gas is able to cool sufficiently, new stars may be formed beyond the extent of the original 
source galaxy, as seen in the simulations of  Kapferer et al. (2009).
Individual examples of intra-cluster star formation have been observed previously in nearby 
and distant clusters (Gerhard et al. 2002; Cortese et al. 2004, 2007; Boquien et al. 2007; Sun 
et al. 2007; Reverte et al. 2007; Yoshida et al. 2008), and our Coma GSE sample 
shows that this process may be widespread.
Unlike the gas from which they formed, the stars do not experience ram pressure from the ICM, and are free to move under gravity,
falling back towards their parent galaxy. 
The ``de-coupling'' of newly-formed stars from the ICM wind leads to a characteristic gradient in the stripped trails. 
While the star-formation event is ongoing, the stripped material will be luminous both in the UV and in H$\alpha$ emission. However, 
because the H$\alpha$ emission fades rapidly after star formation ceases ($\sim$10\,Myr), the streams of back-falling stars will be most prominent
in the UV continuum, for which the timescale is 100--1000\,Myr. This scenario is supported in some individual cases, notably 
GMP 4060 where the ``fireballs'' at the ends of the filaments are luminous in H$\alpha$ but the filaments closer to the galaxy are seen only in UV. 
In some configurations, the newly-formed stars will remain bound to the source galaxy, and may be able to rejoin it, contributing to the
growth of a central bulge, as in the simulations of Kapferer et al. (2009).
Given the low total mass of stars likely being produced in the tails, however, this process seems unlikely to resolve 
the discrepancy between bulge masses in cluster S0s and their supposed spiral progenitors (e.g. Kodama \& Smail 2001; Wilman et al. 2009). 

\begin{figure*}
\includegraphics[angle=0,width=180mm]{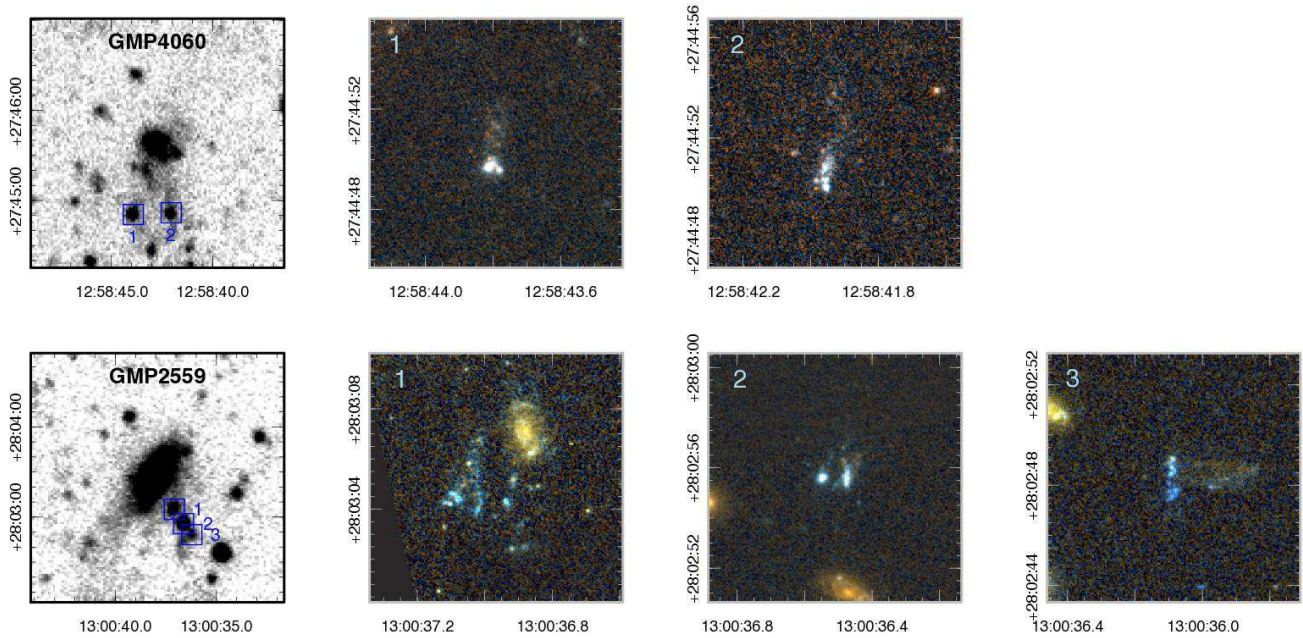}
\caption{Colour ACS images of the ``fireballs'' in GMP 4060 (upper panels) and GMP 2559 (lower panels), based on F814W and F475W imaging from
Carter et al. (2008). The first panel in each row is the {\it GALEX} combined FUV and NUV image, for orientation. The ACS sub-images are 11\,arcsec on a side, and 
correspond to the blue squares in the first panel.}
\label{fig:fireballs}
\end{figure*}

Alternatively, the newly-formed stars may become unbound from the source galaxy, contributing to a genuinely intra-cluster population of stars 
or star clusters. Such intra-cluster star formation has been tentatively identified in recent hydrodynamic simulations of galaxy clusters
by Puchwein et al. (2010), where it contributes some 30\,per cent of the total intra-cluster light. 
If the stars formed in the fireballs were to  remain internally bound in compact systems, these would constitute an unusual class of objects
analogous to the ``tidal dwarf galaxies'' (TDG) identified in some interacting galaxies (e.g. Hancock et al. 2009). 
Two galaxies from our sample (GMP 2559 and GMP 4060) were observed in the HST survey of Carter et al. (2008), providing high-resolution
images of the star-forming clumps in these objects (Figure~\ref{fig:fireballs}).
From the this data, we find that the brightest clumps have luminosities $10^6-10^7$L$_{\odot}$ in the $I$-band. Based on the 
Starburst99 models of Leitherer et al. (1999), their colours of $B-I\approx0.2$ suggest $I$-band mass-to-light ratios of 
$\sim$0.01\,M$_{\odot}$/L$_{\odot}$ (neglecting dust). 
Thus the stellar mass of these structures is probably $\sim10^4-10^5$M$_\odot$, similar to globular clusters.
From their H$\alpha$ fluxes, we estimate the star-formation rates to be 0.001--0.004$\,{M_\odot}$yr$^{-1}$, and hence their star-formation time-scales 
are $<$1.0\,Gyr. The fireball masses are similar to those estimated for the TDGs in Arp 305 by Hancock et al. (2009). 
For TDGs in interacting field galaxies, it is a matter of dispute whether
they will eventually become independent of the galaxies which originally hosted their gas. By contrast, in the case of ram-pressure stripped
cluster galaxies, the ultimate detachment of the clumps from the host galaxy seems quite likely, and if they survive as bound systems they 
may evolve into stellar systems resembling intra-cluster globular clusters or compact dwarf galaxies. Similar objects have been noted in simulations
of ram-pressure stripping by Kapferer et al. (2008), who term them ``stripped baryonic dwarfs''.

\subsection{Post-stripping galaxies and HI deficiency}

Finally, we consider the destiny of the galaxies themselves after stripping is completed. 
It has been known for a long time that cluster spirals are deficient in neutral gas, relative to their counterparts in the field
(e.g. Haynes \& Giovanelli 1984). In Coma, Gavazzi et al. (2006) find that a significant average HI deficiency 
extends from the cluster core out to $\sim2$\,Mpc, the gas content becoming consistent with a field reference sample at $\sim$3\,Mpc.
(A similar trend is seen for Virgo, e.g. Cayatte et al. 1994).
The HI deficiency data for Coma show an apparently sharp transition at a radius of $\sim$1\,Mpc, within which nearly all cluster members are 
gas-poor compared to field spirals. 

Figure~\ref{fig:stripfrac} compares the  Gavazzi et. al. HI-deficient fraction to the incidence of ongoing 
stripping events identified in this paper.  For this test, the galaxies are flagged as gas-deficient if they have Def$_{\rm HI}>0.64$, corresponding to 
the 95th percentile of Def$_{\rm HI}$ among galaxies beyond 3\,Mpc from the Coma core. 
It is notable that a similar characteristic radius of  $\sim$1\,Mpc seems to apply to both phenomena.
On the other hand we found no ongoing stripping events, with the uncertain exception of GMP 5422, beyond 1\,Mpc,
where 20--30\,per cent of spirals are HI-deficient. This result can be understood in terms of a ``backsplash'' population 
(Sanchis et al. 2002; Gill, Knebe \& Gibson 2004): 
although stripping itself is only effective within $\sim$1\,Mpc, HI-deficient {\it post-stripping} galaxies can be 
observed at larger radii after the initial stripping event is complete and the galaxy has passed through the cluster core.
This is confirmed by Figure~\ref{fig:millengse}, which shows that our simple stripping model, tuned to reproduce the fraction of GSE 
galaxies, also produces a post-stripping population consistent with the observed HI-deficient galaxy fraction.

\section{Conclusions}\label{sec:concs}

We have used UV and optical imaging to identify a sample of candidate gaseous stripping events in the Coma cluster. The stripped galaxies are
characterised by tails or trails of UV-bright debris, which we interpret as young stars formed within gas stripped by ram pressure
from the intra-cluster medium. Some of these cases have been noted as peculiar in previous work, in a variety of wavebands
 (Vollmer et al. 2001; Finoguenov et al. 2004; Yagi et al. 2007; Yoshida et al. 2008; Miller et al. 2009), while others
are newly identified here as possible stripping events. 

The trails are predominantly oriented away from the cluster centre, indicating that the galaxies are falling into the cluster for the first time,
along fairly radial orbits, and that the stripping events are completed rapidly compared to the orbital time-scale.  
All but one uncertain case lie at projected radii of 300--900\,kpc from the cluster centre. The radial distribution
of these galaxies is much more centrally concentrated than the distribution of blue galaxies from which they were selected, and more similar
to the distribution of passive galaxies. Within 1\,Mpc projected radius, some 30\,per cent of blue galaxies are currently undergoing stripping,
a fraction which is compatible with a $\sim$500\,Myr time-scale for the stripping events. 
The radius within which UV trails are observed corresponds to an ICM density of $\sim10^{-27}$\,g\,cm$^{-3}$, in agreement with simulations
which show significant star formation in the stripped wake in this density regime for infall velocities $\sim$1000\,\kms\ (Kapferer et al. 2009). 
There are hints that some stripping events are associated with local enhancements in the ICM density, e.g. the western structure and the
NGC 4839 group, but a firm link can not be concluded from the present data. 

We propose an interpretation of these objects as a stage in ram-pressure stripping that is subsequent to the HI gas-tail phase 
(Chung et al. 2007), and occurring at higher ambient densities. The star formation triggered in the stripping events may add mass to the
galaxy bulge, if newly-formed stars fall back into the source galaxy. Alternatively they may escape, forming intra-cluster stellar
systems that could evolve into objects resembling globular clusters or compact dwarf galaxies. After the initial 
stripping, the infalling galaxies will remain as gas-deficient spirals before fading slowly into S0s as they exhaust their remaining gas. 

As stressed by Sun et al. (2010), a fuller understanding of the relationship between different manifestations of gas stripping (HI deficiency, 
and tails in HI, UV, H$\alpha$ and X-ray) will be made possible by improving the overlap between observations in the various wavebands,
for the same galaxy cluster. Our work has assembled a comprehensive wide-field optical, UV, H$\alpha$ and spectroscopic dataset for Coma,
complemented by archival XMM imaging and the radio continuum survey of Miller et al. (2009). A key missing element is high-sensitivity 21cm HI
mapping of a large sample of Coma member galaxies, which should be possible in the next few years using the Expanded Very Large Array.

\section*{Acknowledgments}

We are grateful to Stephen Gwyn for generating a custom stack of the Adami deep $u$-band data for our use, 
to Masafumi Yagi for communicating the Subaru H$\alpha$ results in advance of submission, and to Neal Miller for helpful comments on this paper. 
RJS was supported for this work by 
STFC Rolling Grant PP/C501568/1 ``Extragalactic Astronomy and Cosmology at Durham 2008--2013''.
This work is based on observations made with the NASA {\it Galaxy Evolution Explorer (GALEX)}. {\it GALEX} is a NASA Small Explorer, developed in 
cooperation with the Centre National d'Etudes Spatiales of France and the Korean Ministry of Science and Technology. 
This work is based on observations obtained with MegaPrime/MegaCam, a joint project of CFHT and CEA/DAPNIA, at the Canada--France--Hawaii Telescope (CFHT) which is operated by the National Research Council (NRC) of Canada, the Institute National des Sciences de l'Univers of the Centre National de la Recherche Scientifique of France, and the University of Hawaii. The work has made use of data products produced at the TERAPIX data center located at the Institut d'Astrophysique de Paris.
The Isaac Newton Telescope is operated on the island of La Palma by the Isaac Newton Group in the Spanish Observatorio del Roque de los Muchachos of
the Instituto de Astrof\'isica de Canarias.
This work has made use of the NASA/IPAC Extragalactic Database (NED) which is operated by the Jet Propulsion Laboratory, California Institute of Technology, under contract with the National Aeronautics and Space Administration.
The Millennium Simulation databases used in this paper and the web application providing online access to them were constructed as part of the activities of the German Astrophysical Virtual Observatory.

{}

\label{lastpage}
\end{document}